\documentclass[aps,preprintnumbers, superscriptaddress, amsmath, showpacs]{revtex4}
\usepackage{psfrag}
\usepackage{epsfig}
\usepackage{amsfonts} 
\usepackage{amssymb}  
\usepackage{graphicx} 
\usepackage{amsmath}  
\usepackage{amsmath}  

\usepackage[latin1]{inputenc}
\usepackage{color}
\usepackage{suffix}
\usepackage{mathtools}

\newcommand{\slsh}[1]{{\not \! #1}}
\newcommand{\Eq}[1]{{Eq.~({\ref{#1}})}}
\newcommand{\bea}{\begin{eqnarray}}
\newcommand{\eea}{\end{eqnarray}}
\newcommand{\beas}{\begin{eqnarray*}}
\newcommand{\eeas}{\end{eqnarray*}}
\newcommand{\sumint}{\sum\!\!\!\!\!\!\!\!\int}

\DeclarePairedDelimiterX\MeijerM[3]{\lparen}{\rparen}%
{\begin{smallmatrix}#1 \\ #2\end{smallmatrix}\delimsize\vert\,#3}

\newcommand\MeijerG[8][]{%
  G^{\,#2,#3}_{#4,#5}\MeijerM[#1]{#6}{#7}{#8}}

\WithSuffix\newcommand\MeijerG*[7]{%
  G^{\,#1,#2}_{#3,#4}\MeijerM*{#5}{#6}{#7}}

\def\Title#1{\begin{center} {\Large {\bf #1} } \end{center}}

\begin{document}

\Title{Magnetic Field Effect in the Fine-Structure Constant and Electron Dynamical Mass}

\author{E. J. Ferrer}
\affiliation{Dept. of Physics and Astronomy, Univ. of Texas Rio Grande Valley, 1201 West University Dr., Edinburg, TX 78539 and CUNY-Graduate Center, New York 10314, USA}
 \author{A. Sanchez}
\affiliation{Facultad de Ciencias, Universidad Nacional Aut\' onoma de M\' exico, Apartado Postal 50-542, Ciudad de M\'exico  04510, Mexico.}

\begin{abstract}

We investigate the effect of an applied constant and uniform magnetic field in the fine-structure constant of massive and massless QED. In massive QED, it is shown that a strong magnetic field removes the so called Landau pole and that the fine-structure constant becomes anisotropic having different values along and transverse to the field direction. Contrary to other results in the literature, we find that the anisotropic fine-structure constant always decreases with the field. We also study the effect of the running of the coupling constant with the magnetic field on the electron mass. We find that in both cases of massive and massless QED, the electron dynamical mass always decreases with the magnetic field, what can be interpreted as an inverse magnetic catalysis effect.
\end{abstract}


\maketitle

\section{Introduction}

The effect of magnetic fields on the different properties of quantum particles has always attracted great interest \cite{Rafelski}. At present, it has been reinforced by the fact that there is the capability to generate very strong magnetic fields in non-central heavy ion collisions, and because of the discovery of strongly magnetized compact stars, which have been named magnetars. In this domains, fields of the order or larger than the QCD scale ($B \geqslant \Lambda^2_{QCD}=4\times10^4 MeV^2=6.8\times10^{18}$ G) are estimated to be generated at RHIC and LHC \cite{HIC}, and fields larger than the electron critical field, $B_c^{(e)}=4.4\times 10^{13}$ G, can exist in the surface of magnetars \cite{magnetars}.

Among the magnetic field effects studied in electrodynamics, those related to the possible variation of the fine-structure constant  \cite{Ragazzon-Alfa}-\cite{Chodos-2} and to the electron mass \cite{Jancovici-1969}-\cite{Paraelectricity}  have attracted special attention. 

In QED, screening effects can modify the value of the observable coupling constant. There exists even the possibility that the renormalized coupling becomes screened to zero. In that case, the theory is said to be "trivial". Thus, a theory that appears to describe interacting particles at the classical level, can become a "trivial" theory of noninteracting free particles when quantum and relativistic effects are included. This phenomenon is referred to as quantum triviality \cite{Triviality}. 
This problem appears in QED in what is known as the Landau pole problem \cite{Landau-pole}, where QED becomes inconsistent at very short-distance scales in the perturbative regime unless the renormalized charge is set to zero. The inclusion of a magnetic field can affect this result since the magnetic field contribute to the charge screening. In this paper, we will show that in a strong magnetic field the Landau-pole is removed by the screening produced by the electron-positron pairs filling the lowest Landau level (LLL). On the other hand, we obtain that the behavior of the fine-structure constant with the magnetic field is in disagreement with those reported in Refs. \cite{Ragazzon-Alfa, Chodos} years ago. We will make a detailed exposition of the gauge invariant method we are using so to show what is the source of this discrepancy.

Another problem that has attracted much attention in the last two decades is the so-called magnetic catalysis of chiral symmetry breaking (MC$\chi$SB) \cite{MC}-\cite{Paraelectricity}. This phenomenon is responsible for the dynamical generation of a fermion mass (i.e. by catalyzing chiral symmetry breaking) by an applied magnetic field in a massless fermion theory.  The study of theories of massless relativistic fermions has recently gained new interest in the context of quasiplanar systems, such as pyrolitic graphites (HOPG) \cite{3, 4} and graphene \cite{5}, because their low-energy excitation quasiparticle spectrum has a linear dispersion relation. The dynamics of those charge carriers is described by a relativistic quantum field theory of massless fermions in 2+1 dimensions \cite{3, 6}.
The MC$\chi$SB is a universal phenomenon that takes place in any relativistic theory of interactive massless fermions in a magnetic field, and it has been proposed as the mechanism explaining various effects in quasiplanar condensed matter systems \cite{7}.

In the original works of MC$\chi$SB a characteristic feature is the increase of the dynamical mass with the magnetic field (for a review see \cite{MC-Reviews} and references therein). Hence, as the critical temperature for chiral restoration results proportional to the dynamical mass, the critical temperature also increases with the field  \cite{Gusynin}. Nevertheless, this result is in sharp contrast with recent QCD-lattice calculations that showed a decrease of the critical temperature for the chiral/deconfinement transition with the magnetic field \cite{Inverse-MC}, a phenomenon that has been called "inverse magnetic catalysis" (IMC).

In reference \cite{Wen}, we adopted the point of view, shared by other authors \cite{IMC-running-const}, that the origin of the IMC in QCD should lie in the effects of the magnetic field in the running of the strong coupling. Our analysis contained two new fundamental elements. On the one hand, we showed that in the strong field region ($qB> \Lambda^2_{QCD}$), where the infrared dynamics becomes relevant, the QCD running coupling becomes anisotropic: the color interaction in the directions parallel and transverse to the field is characterized by two different functions of the momentum and the field \cite{Wen}. On the other hand, we found that the quarks, confined by the field to the LLL, produce magnetic antiscreening (i.e. the quark magnetic contribution to the running coupling constant enters with the same sign as the gluon contribution) in the parallel coupling, which is the one entering to determine the chiral critical temperature. The magnetic antiscreening of the LLL quarks is connected to the color paramagnetic behavior of the pairs formed by LLL virtual quarks and antiquarks \cite{Nielsen}. The magnetic antiscreening produced by the LLL pairs increases with the magnetic field because the phase space of the LLL increases with the field, allowing more pairs to be formed. 
These results naturally lead to IMC and also allow us to identify a possible physical mechanism for the behavior of Tc with the field (i.e. the decrease of the parallel strong coupling with the field).

From this result and taking into account the universal character of the MC$\chi$SB phenomenon, it is natural to ask if  the IMC effect also takes place in massless QED. To get inside on this question will be another goal of this paper. As we will see below, there are some signals of IMC in this case, although, as it will be discussed there, more work is needed for a complete certain answer. From a physical point of view, the question is to find if the weakening effect on the coupling produced by the screening of the pairs into the LLL, is surpassed  by the strengthen of the interaction due to the reduction of the spatial dimension produced by the particle confinement to the LLL. We will show that in this case, as in the QCD case, the weakening of the coupling constant is the winning effect. 

The paper is organized as follows: In Section II, we calculate the Coulomb potential energy in the presence of a constant and uniform magnetic field considering one-loop corrections through the polarization operator in the two limits of strong and weak magnetic fields. In doing this, for the sake of completeness, we review the approach introduced in Ref. \cite{Shabad} making more explicit some derivations. Then, in Section III, we use the results of Section II to calculate the running of the fine-structure constant with the magnetic field at different strengths. 
In Section IV, we investigate how the magnetic field affects the electron mass in massive QED, as well as in the massless case where the MC$\chi$SB phenomenon plays a fundamental role. In this analysis, an important new element is that the effect of the magnetic field in the fine-structure constant for each case is included. Then, the possibility of IMC in the massless case is analyzed. In Section V, the main results of this paper are summarized and their physical significance are discussed. Finally, in Appendix A, detailed calculations of the polarization-operator coefficient entering in the Coulomb potential energy are given in the weak and strong-field approximations using the Ritus's method.

\section{The Coulomb potential energy at $B\neq0$}\label{section2}

One of the main goals in this paper is to find how a magnetic field can affect the electron mass through radiative and nonperturbative  corrections when the effect of the magnetic field is also considered in the fine-structure constant. To find how $\alpha_{QED}$ depends on B, we will start by calculating the Coulomb potential energy in different field-strength limits. For the sake of understanding, we review as follows the basic derivations introduced in  Ref. \cite{Shabad}  to study the Coulomb potential in the presence of a magnetic field. To have this explicit derivations will serve then to make it clear what is the source of the discrepancies with previous results regarding the behavior of the fine-structure constant with the magnetic field  \cite{Ragazzon-Alfa, Chodos}. In particular, we will show that the discrepancy is due to the fact that in  \cite{Ragazzon-Alfa, Chodos} it was considered in the structure of the photon propagator, terms that are forbidden by the gauge invariance of the polarization operator. 

From the relationship between the 4-vector potential $A_\mu(x)$ and the 4-current $J_\mu(y)$,
\begin{equation}\label{4-potential}
A_\mu(x)=\int_{-\infty}^\infty D_{\mu \nu}(x-y)J^\nu(y)d^4y,
\end{equation}
where $D_{\mu \nu}(x-y)$ is the photon propagator, we have that for a static point-charge source, $J^\nu(y)=e\delta_{\nu 0} \delta^3(\textbf{y})$, the corresponding static potential is 
\begin{equation}\label{static-potential}
A_\mu(\textbf{x})=e\int_{-\infty}^\infty D_{\mu 0}(y_0, \textbf{x})dy_0.
\end{equation}

Since the photon fields are electrically neutral, the photon propagator can be Fourier transformed 
\begin{equation}\label{Fourier-Trans}
D_{\mu \nu}(x)=\frac{1}{(2\pi)^4} \int D_{\mu \nu}(k) e^{ikx}d^4k
\end{equation}

Then, substituting (\ref{Fourier-Trans}) in (\ref{static-potential}), we obtain the general expression for the Coulomb potential of a point-like static charge,
\begin{equation}\label{Coulomb-Pot}
A_0(\textbf{x})=\frac{e}{(2\pi)^3}\int D_{00}(0,\textbf{k})e^{-i\textbf{k}\cdot \textbf{x}}d^3k
\end{equation}

If a second point-charge $e$ is located at $ \textbf{x}$, the system electrostatic energy is
\begin{equation}\label{Coulomb-Pot-Energy}
U(x)=\frac{e^2}{(2\pi)^3}\int D_{00}(0,\textbf{k})e^{-i\textbf{k}\cdot \textbf{x}} d^3k
\end{equation}

Therefore, to find the Coulomb electrostatic energy in the presence of a uniform and constant magnetic field taken in the Landau gauge $A_\mu=(0,0,Bx_1,0)$, we need to know the photon propagator. With that goal, we introduce the four orthogonal vectors, which can expand the 4-dimensional space in the presence of a magnetic field \cite{Paraelectricity, Shabad, Batalin}
\begin{equation}\label{Egin-Vectors}
b_\mu^{(1)}=-k^2\hat{F}_{\mu \nu} \hat{F}_{\nu \rho} k_\rho+k_\bot^2 k_\mu, \quad b_\mu^{(2)}=\frac{1}{2}\epsilon_{\mu \nu \rho \lambda} \hat{F}_{\nu \rho}k_\lambda, \quad b^{(3)}_\mu=\hat{F}_{\mu \nu} k_\nu, \quad b_\mu^{(4)}=k_\mu,
\end{equation}
with $\hat{F}_{\mu \nu}=F_{\mu \nu}/B$, denoting the normalized electromagnetic strength tensor and $k_\bot^2=k_1^2+k_2^2$.  

The four-vectors (\ref{Egin-Vectors}) satisfy the relations
\begin{equation}\label{rel-1}
\sum_{a=1}^3 \frac{b_\mu^{(a)}b_\nu^{(a)}}{(b^{(a)})^2}=g_{\mu \nu}-\frac{k_\mu k_\nu}{k^2},
\end{equation}
\begin{equation}\label{rel-2}
\sum_{a=1}^4 \frac{b_\mu^{(a)}b_\nu^{(a)}}{(b^{(a)})^2}=g_{\mu \nu}
\end{equation}

The polarization tensor is a second-rank Lorentz tensor transverse to $k_\mu$ (i.e. gauge invariant), which is also P and C invariant. As a consequence, it is diagonal and can be given in momentum space in terms of the orthogonal vectors  (\ref{Egin-Vectors}) as \cite{Batalin, Shabad-Trudy},
\begin{equation}\label{Polarization-Opr}
\Pi_{\mu \nu}(k, B)=\sum_{a=1}^3\Pi_a(k,B)\frac{b_\mu^{(a)} b_\nu^{(a)}}{(b^{(a)})^2}
\end{equation}
The coefficients $\Pi_a(k,B)$ are scalars that depend on the momentum and magnetic field. As follows from (\ref{Polarization-Opr}), and the orthogonality of the eigenvectors $b_\mu^{(a)}$, the four-vectors (\ref{Egin-Vectors}) are the eigenvectors of the polarization operator with corresponding eigenvectors $\Pi_a(k,B)$,
\begin{equation}
\Pi_\mu^\nu(k,B)b_\nu^{(a)}=\Pi_a(k,B)b_\mu^{(a)} 
\end{equation}

The inverse propagator for the Maxwell theory in a covariant gauge and including the radiative correction associated with the polarization operator is then given by
\begin{equation}\label{Inv-Prog-1}
D_{\mu \nu}^{-1}(k,B)=-k^2g_{\mu \nu} +(1-\frac{1}{\xi})k_\mu k_\nu+\Pi_{\mu \nu}(k,B),
\end{equation} 
where $\xi$ is the gauge-fixing parameter. Taking into account the relations (\ref{rel-1}) and (\ref{Polarization-Opr}), the inverse propagator can be written as
\begin{equation}\label{Inv-Prog-2}
D_{\mu \nu}^{-1}(k,B)=\sum_{a=1}^4D^{-1}_a(k,B)\frac{b_\mu^{(a)} b_\nu^{(a)}}{(b^{(a)})^2},
\end{equation} 
with coefficients
\begin{equation}
D^{-1}_a(k,B) =
\begin{cases} 
[\Pi_a(k,B)-k^2], &  a={1,2,3}, 
\\
\quad k^2/\xi, & a=4.
\end{cases}
\label{D}
\end{equation}

By using the orthogonality conditions of the eigenvectors (\ref{Egin-Vectors}) and the relation (\ref{rel-2}), the photon propagator including the radiative corrections can be easily found from
\begin{equation}
D_{\mu \rho}^{-1}(k,B)D^{\rho \nu}(k,B)=\delta_\mu^\nu,
\end{equation}
 to be given as
\begin{equation}\label{Propagator-B}
D_{\mu \nu}(k,B)=\sum_{a=1}^4D_a(k,B)\frac{b_\mu^{(a)} b_\nu^{(a)}}{(b^{(a)})^2},
\end{equation}
where  
\begin{equation}
D_a(k,B) =
\begin{cases} 
[\Pi_a(k,B)-k^2]^{-1}, &  a={1,2,3}, 
\\
\quad \xi/k^2, & a=4.
\end{cases}
\label{D-1}
\end{equation}

It can be noticed that in the Feynman gauge, $\xi=0$, the tensor structure of the photon propagator (\ref{Propagator-B}) reduces to that of the polarization operator (\ref{Polarization-Opr}).

Hence, from (\ref{Coulomb-Pot-Energy}), (\ref{D}) and  (\ref{Propagator-B}) we have,
\begin{eqnarray}\label{Coulomb-Pot-2}
U(x)=\frac{e^2}{(2\pi)^3}\int \frac {e^{-i\textbf{k}\cdot \textbf{x}}d^3k}{ [\textbf{k}^2-\Pi_2(0,k_3^2,k_\bot^2)]} \qquad
\end{eqnarray}

Since only the contribution $a=2$ in (\ref{D-1})  is different from zero for $D_{00}(0,\textbf{k)}$, we see that the Coulomb potential energy is gauge independent (i.e. it does not depend on $\xi$).

The coefficient of the polarization operator entering in (\ref{Coulomb-Pot-2}), $\Pi_2(0, \textbf{k},B)$, depends on the approximation. In the following, we will consider the polarization operator in the one-loop approximation taken in the two extreme values of the magnetic field.

\subsection{Strong-Field Approximation}

The coefficient  $\Pi_2(0, \textbf{k},B)$, in the strong-field approximation ($eB\gg m^2,k^2$), was originally calculated in \cite{Batalin, Shabad-Trudy} to be given in the leading approximation by
\begin{equation}\label{Coefficient-2}
 \Pi_2^{(s)}(k_0=0,m^2<k^2_\bot,  k^2_3 < eB)=-\frac{2\alpha_0 |eB|}{\pi}e^\frac{-k_\bot^2}{2|eB|}, 
 \end{equation}

\begin{equation}\label{Coefficient-3}
 \Pi_2^{(s)}(k_0=0,k^2_\bot,  k^2_3<m^2 < eB)=-\frac{\alpha_0 k_3^2 |eB|}{3\pi m^2}e^\frac{-k_\bot^2}{2|eB|}
\end{equation}

It is important to stress that these coefficients were calculated in Refs. \cite{Batalin, Shabad-Trudy} using the Schwinger proper-time approach \cite{Proper-Time}, where the contribution of the different Landau levels is not apparent. Nevertheless, if we use the Ritus's approach \cite{Ritus:1978cj}, where the Landau level contribution becomes explicit,  we can show (see Appendix \ref{strong-app}) that the strong-field values (\ref{Coefficient-2}) and (\ref{Coefficient-3}) are obtained directly working in the LLL limit. This is an evidence that the fermions contributing to the Coulomb potential energy in this limit are confined to (1+1)-dimensions. 

Taking into account the results (\ref{Coefficient-2}) and (\ref{Coefficient-3}), we obtain respectively for the Coulomb potential energy (\ref{Coulomb-Pot-2}) in the LLL approximation
\begin{equation}\label{Pot-strong-1}
U^{(s)}(k)=\frac{\alpha_0} {\textbf{k}^2\left[1+\frac{2\alpha_0|eB|}{\pi \textbf{k}^2}e^\frac{-k_\bot^2}{2|eB|} \right] }, \quad m^2<k^2_\bot,  k^2_3 < eB,
\end{equation}
and 
\begin{equation}\label{Pot-strong-2}
U^{(s)}(k)=\frac{\alpha_0} {\textbf{k}^2\left[1+\frac{\alpha_0k_3^2|eB|}{3\pi m^2 \textbf{k}^2}e^\frac{-k_\bot^2}{2|eB|} \right] }, \quad k^2_\bot,  k^2_3<m^2 < eB,
\end{equation}

\subsection{Weak-Field Approximation}

In the weak-field approximation ($eB<m^2, k^2$) the coefficient $\Pi_2$ if found in Appendix  \ref{strong-app} in two different regions (see Eqs, (\ref{Coefficien-WA-1}) and (\ref{Coefficien-WA-1-2})),
\begin{equation}\label{Coefficien-WA-12}
 \Pi_2^{(w)}(eB<k_\|^2, k_\bot^2<m^2)=-\frac{1}{2}\left[\frac{\alpha_0}{3\pi}\left(\frac{eB}{m^2}\right)^2\left(\frac{k_\bot^2}{5}+\frac{k_\|^2}{2}\right)-\frac{8\alpha_0}{15\pi}\frac{k^4}{m^2}\right]
 \end{equation}

\begin{equation}\label{Coefficien-WA-1-22}
 \Pi_2^{(w)}(eB<m^2<k_\|^2, k_\bot^2)=-\frac{1}{2}\left[\frac{\alpha_0}{3\pi}\left(\frac{eB}{m^2}\right)^2\left(\frac{k_\bot^2}{5}+\frac{k_\|^2}{2}\right)-\frac{4\alpha_0}{3\pi}k^2\ln\left(\frac{k}{m}\right)\right]
\end{equation}

Then, as in the strong-field case, from (\ref{Coefficien-WA-12}) and (\ref{Coefficien-WA-1-22}) we obtain respectively for the Coulomb potential energy (\ref{Coulomb-Pot-2}) in the weak-field approximation
\begin{equation}\label{Pot-Weak-1}
U^{(w)}(k)=\frac{\alpha_0} {\textbf{k}^2\left[1+\frac{\alpha_0}{6\pi\textbf{k}^2}\left(\frac{eB}{m^2}\right)^2\left(\frac{k_\bot^2}{5}+\frac{k_3^2}{2}\right)-\frac{4\alpha_0}{15\pi}\frac{\textbf{k}^2}{m^2} \right] }, \quad eB<k^2_\bot,  k^2_3 < m^2,
\end{equation}
and 
\begin{equation}\label{Pot-weak-2}
U^{(w)}(k)=\frac{\alpha_0} {\textbf{k}^2\left[1+\frac{\alpha_0}{6\pi\textbf{k}^2}\left(\frac{eB}{m^2}\right)^2\left(\frac{k_\bot^2}{5}+\frac{k_3^2}{2}\right)-\frac{2\alpha_0}{3\pi}\ln\left(\frac{k}{m}\right) \right] }, \quad eB<m^2<k^2_\bot,  k^2_3
\end{equation}

\section{The running of the fine-structure constant with $B$}\label{section3}

The field dependent fine-structure constant, $\alpha_{QED}$, can be obtained from the Coulomb potential energy (\ref{Coulomb-Pot-2}) written as
\begin{eqnarray}\label{Coulomb-Pot-3}
U(x)=\frac{1}{2\pi^2}\int \frac {\alpha_0}{\textbf{k}^2 \left[1-\frac{1}{\textbf{k}^2} \Pi_2(0,k_3^2,k_\bot^2)\right]}  e^{-i\textbf{k}\cdot \textbf{x}}d^3\textbf{k}=\frac{1}{2\pi^2}\int   \frac {\alpha_{QED}(\textbf{k,B})} {\textbf{k}^2}   e^{-i\textbf{k}\cdot \textbf{x}}d^3\textbf{k}
\end{eqnarray}

Taking into account that the coefficient $\Pi_2(k,B)$ has two different asymptotic behaviors, i.e. at strong and weak magnetic fields respectively, as follows we find the fine-structure constant in those two limits.

\subsection{Weak-Field Limit}

In the weak-field limit the fine-structure constant can be obtained from (\ref{Pot-Weak-1}) and (\ref{Pot-weak-2}) respectively as
\begin{equation}\label{Alfa-weak-1}
\alpha^{(w)}_{QED}(\textbf{k,B})=\frac{\alpha_0} {1+\frac{\alpha_0}{6\pi\textbf{k}^2}\left(\frac{|eB|}{m^2}\right)^2\left(\frac{k_\bot^2}{5}+\frac{k_3^2}{2}\right)-\frac{4\alpha_0}{15\pi}\left(\frac{\textbf{k}^2}{m^2} \right) }, \quad |eB|<k^2_\bot,  k^2_3 < m^2,
\end{equation}
and 
\begin{equation}\label{Alfa-weak-2}
\alpha^{(w)}_{QED}(\textbf{k,B})=\frac{\alpha_0} {1+\frac{\alpha_0}{6\pi\textbf{k}^2}\left(\frac{|eB|}{m^2}\right)^2\left(\frac{k_\bot^2}{5}+\frac{k_3^2}{2}\right)-\frac{\alpha_0}{3\pi}\ln\left(\frac{k^2}{m^2}\right)}, \quad  |eB|<m^2<k^2_\bot,  k^2_3
\end{equation}

In expression (\ref{Alfa-weak-2}), we can notice the existence of the so called Landau pole \cite{Landau-pole}. That is, the negative logarithmic term in the denominator can produce for large enough momentum a pole that rises the coupling constant to an infinite value. Since this result is obtained through perturbative one-loop calculations, it is indicating that the pole is merely a sign of the breaking of the perturbative approximation at strong coupling. Going beyond perturbative calculations with Lattice gauge theory it was obtained that the QED charge at $B\neq 0$ is completely screened for an infinite cutoff \cite{Lattice-Landau-pole}. We call attention that in (\ref{Alfa-weak-2}) the magnetic field contribution enters with opposite sign to the logarithmic momentum dependent term. Thus, its effect is to counteract the divergency, although in this approximation ($|eB|\ll k^2$) it is not enough to avoid the pole.

\begin{figure}
\begin{center}
\begin{tabular}{ccc}
  \includegraphics[width=7cm]{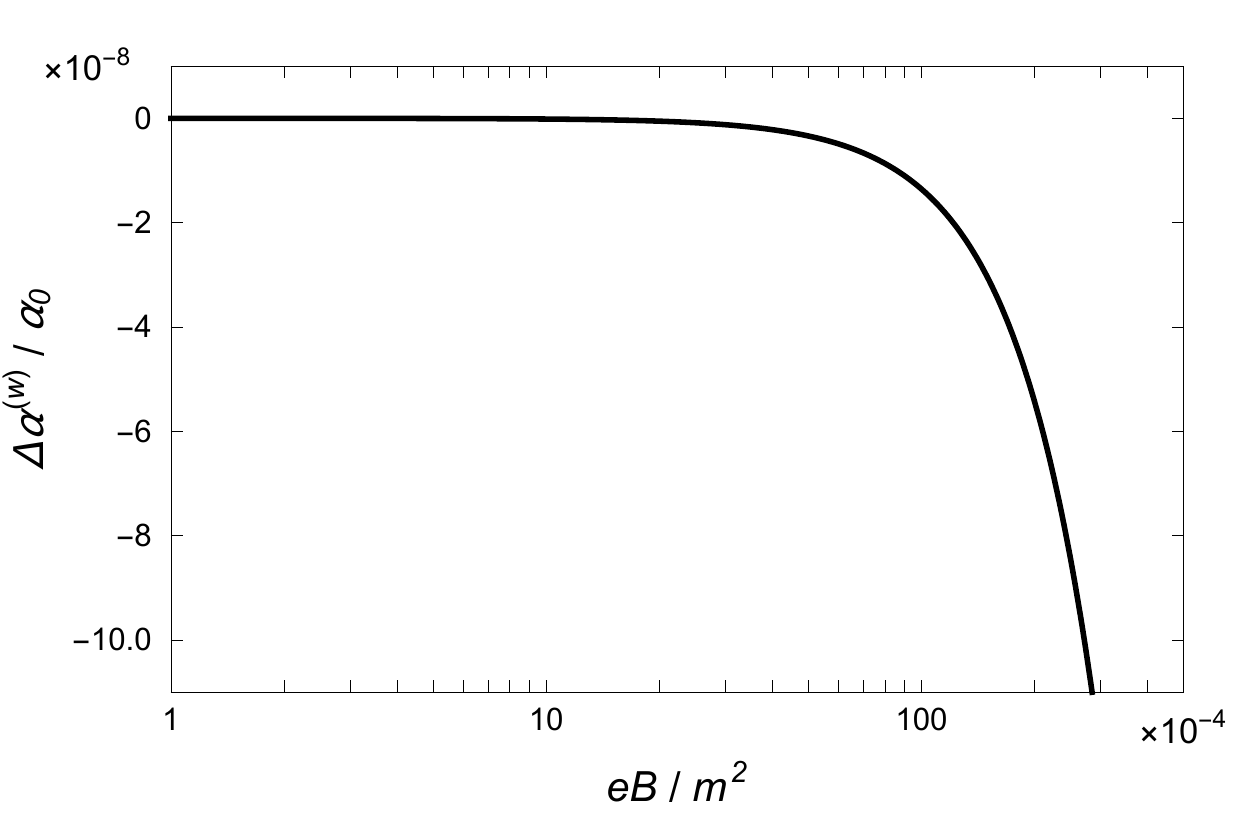} & \includegraphics[width=7cm]{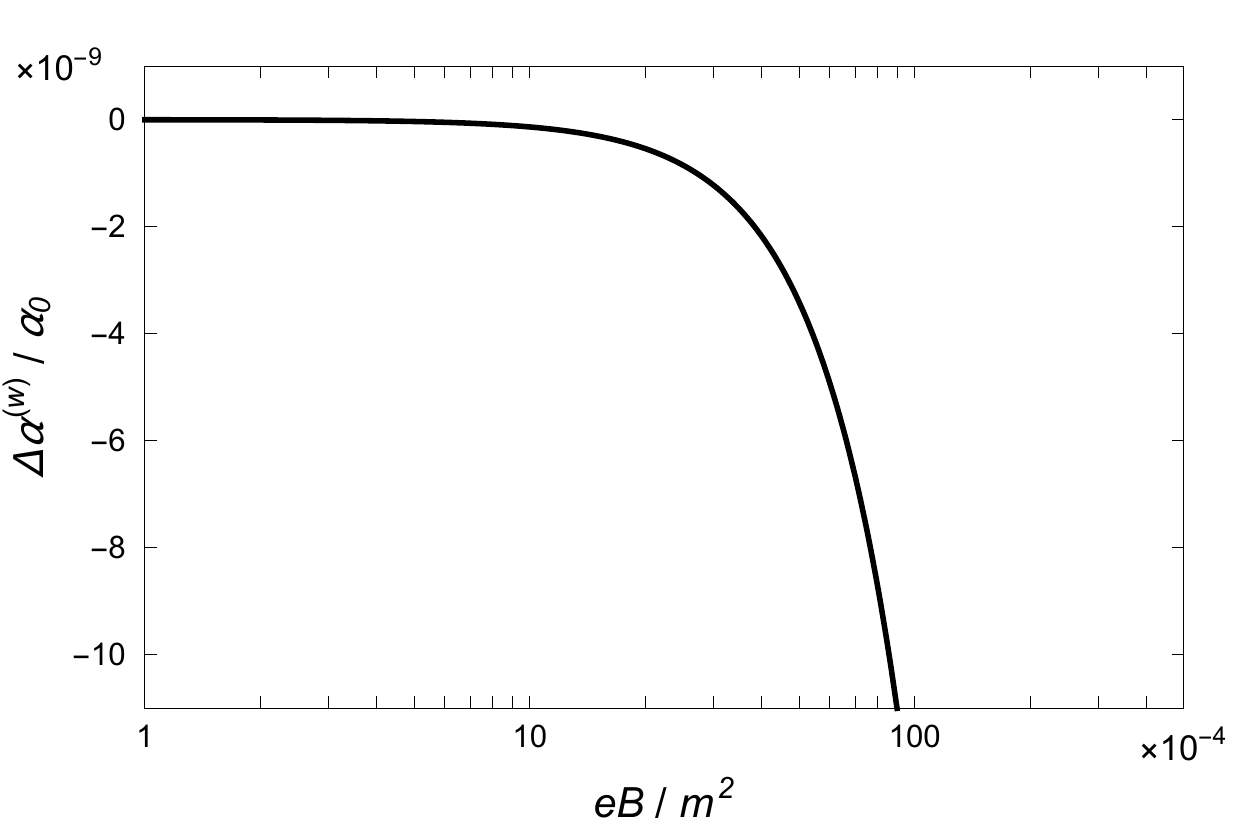}\\
(a) & (b)  \\
  \\
  \end{tabular}
    \end{center}
    \caption{In these figures we plot the relative change of $\alpha^{(w)}(k,B)$ (see Eq.(\ref{deltaalphaw})) versus the magnetic field. (a) Coupling constant given by Eq. (\ref{Alfa-weak-1}) as a function of $eB/m^2$ for fixed values of $k_{||}/m=k_\perp/m=0.5$ and $\alpha_0=1/137$. (b) Coupling constant given by Eq. (\ref{Alfa-weak-2}) as a function of $eB/m^2$ for fixed values of $m/k_{||}=m/k_\perp=0.5$ and $\alpha_0=1/137$. }
     \label{Fig:couplingweaklimit}
\end{figure}

\begin{figure}
\begin{center}
\begin{tabular}{ccc}
  \includegraphics[width=7cm]{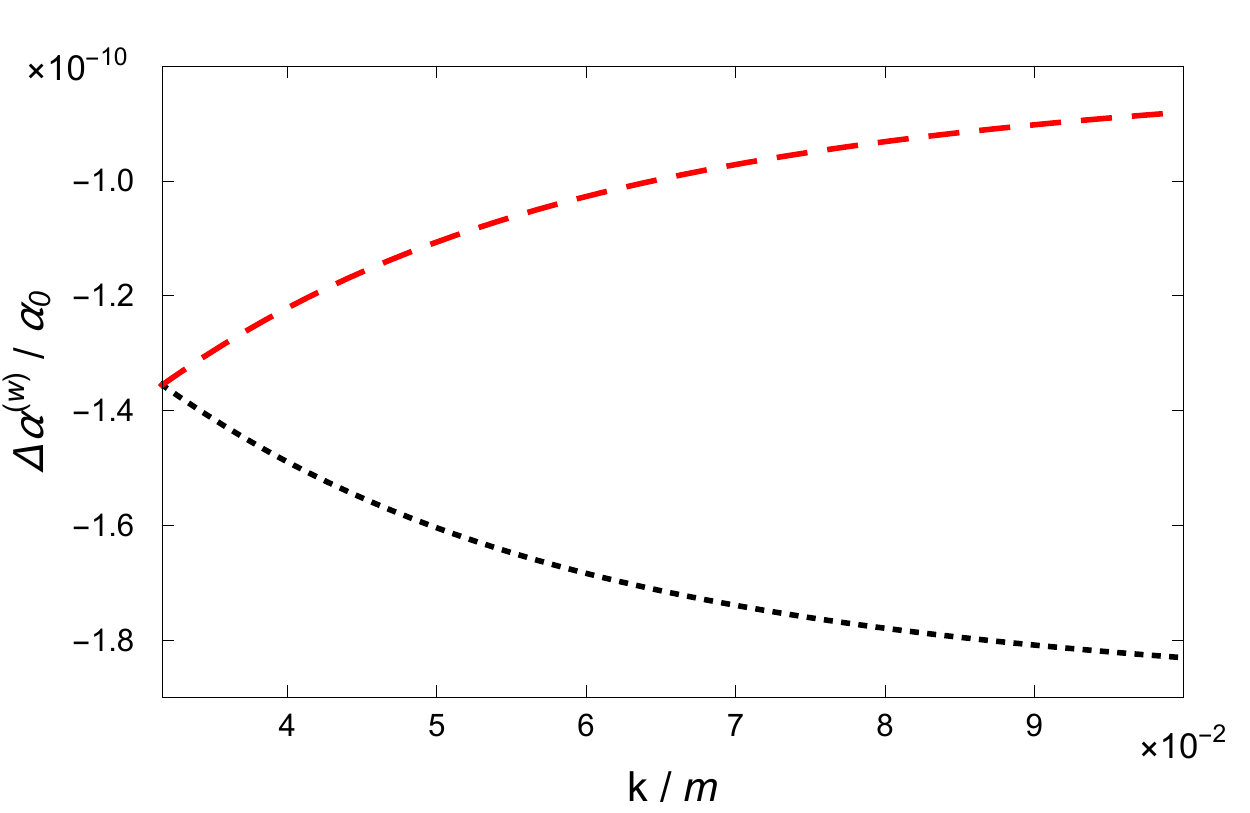} & \includegraphics[width=7cm]{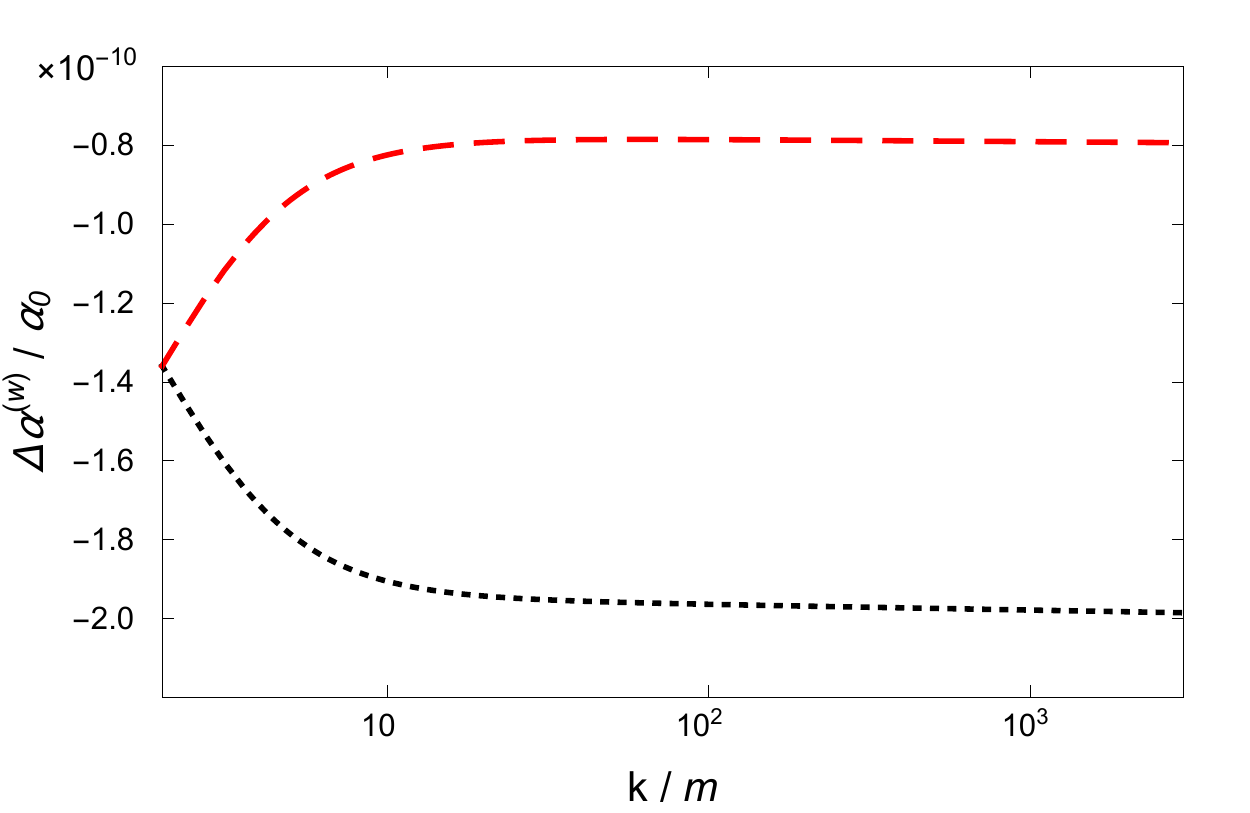}\\
(a) & (b)  \\
  \end{tabular}
    \end{center}
    \caption{(a) Coupling constant given by Eq. (\ref{Alfa-weak-1})  for  $eB/m^2=10^{-3}$ and $\alpha_0=1/137$. {\it Black dotted line}:  Shows the relative change of the coupling constant as a function of $k_{||}/m$ for $k_\perp/m=0.03$. {\it Red dashed line}:  Shows the relative change of the coupling constant as a function of $k_\perp/m$ for $k_{||}/m=0.03$.    (b) Coupling constant given by Eq. (\ref{Alfa-weak-2})  for  $eB/m^2=10^{-3}$ and $\alpha_0=1/137$. {\it Black dotted line}:  Shows the relative change of the coupling constant as a function of $k_{||}/m$ for $k_\perp/m=2$. {\it Red dashed line}:  Shows the relative change of the coupling constant as a function of $k_\perp/m$ for $k_{||}/m=2$.}
\label{Fig:couplingweaklimit-paraperp-1}
\end{figure}

To quantify the magnetic field effect on the coupling constant, let us introduce the relative change of $\alpha^{(w)}(k,B)$ with respect to $\alpha^{(w)}(k,0)$ as
\begin{equation}
       \Delta \alpha^{(w)}\equiv \alpha^{(w)}(k,B)-\alpha^{(w)}(k,0).
\label{deltaalphaw}
\end{equation}
As the plots of Fig. \ref{Fig:couplingweaklimit} and Fig. \ref{Fig:couplingweaklimit-paraperp-1} show, the coupling constant in all cases decreases with the magnetic field and  parallel momentum, while increases with the transverse momentum. Thus, we find that the behavior of the fine-structure constant with the magnetic field is similar to that of the strong coupling constant as reported in  \cite{Wen}.  We can also observe that, in the weak-field approximation, there is a small anisotropy with respect to the directions along and transverse to the magnetic field.

 \subsection{Strong-Field Limit}

From (\ref{Pot-strong-1}) and (\ref{Pot-strong-2}), the coupling constant in the strong-field limit becomes respectively
\begin{equation}\label{Alfa-strong-1}
\alpha^{(s)}_{QED}(\textbf{k},B)\simeq\frac{\alpha_0} {1+\frac{2\alpha_0|eB|}{\pi \textbf{k}^2}e^\frac{-k_\bot^2}{2|eB|}}, \quad m^2<k^2_\bot,  k^2_3 < |eB|,
\end{equation}
and 
\begin{equation}\label{Alfa-strong-2}
\alpha^{(s)}_{QED}(\textbf{k},B)\simeq\frac{\alpha_0} {1+\frac{\alpha_0k_3^2|eB|}{3\pi m^2 \textbf{k}^2}e^\frac{-k_\bot^2}{2|eB|}}, \quad k^2_\bot,  k^2_3<m^2 < |eB|,
\end{equation}

We should notice that the Landau pole \cite{Landau-pole}, that appears for $k^2\gg m^2$ in the weak-field case (\ref{Alfa-weak-2}), is absent in the strong field result (\ref{Alfa-strong-1}). Although the coupling constant continues increasing with the momentum (no-asymptotic free theory) the strong effect of the magnetic field removes the singularity. As we pointed out above, at zero magnetic field the Landau pole is removed in a non-perturbative approach \cite{Lattice-Landau-pole}. Nevertheless, although we are working in the one-loop approximation here, a non-perturbative expansion in the magnetic field is present in the strong-field limit  (\ref{Alfa-strong-1}).  The non-perturbative magnetic-field interaction 
results sufficient to produce a finite coupling constant for all values of the momentum smaller than the natural scale in the strong-field limit $\sqrt{|eB|}$. 

The fine-structure constant in the first limit (\ref {Alfa-strong-1}) decreases with the magnetic field at a fixed momentum. While, in the second limit (\ref{Alfa-strong-2}), we can see that $\alpha^{(s)}_{QED}$ exhibits a significant anisotropy in the directions parallel and transverse to the magnetic field,
\begin{equation}\label{Alfa-strong-perpendicular}
\alpha^{(s)}_{QED}(k_3=0,B)_\bot\simeq\alpha_0, \quad k^2_\bot<m^2 < |eB|,
\end{equation}

\begin{equation}\label{Alfa-strong-parallel}
\alpha^{(s)}_{QED}(k_\bot=0,B)_\|\simeq\frac{\alpha_0} {1+\frac{\alpha_0|eB|}{3\pi m^2}},   \quad k^2_\|<m^2 < |eB|,
\end{equation}

\begin{figure}
\begin{center}
  \includegraphics[width=8cm]{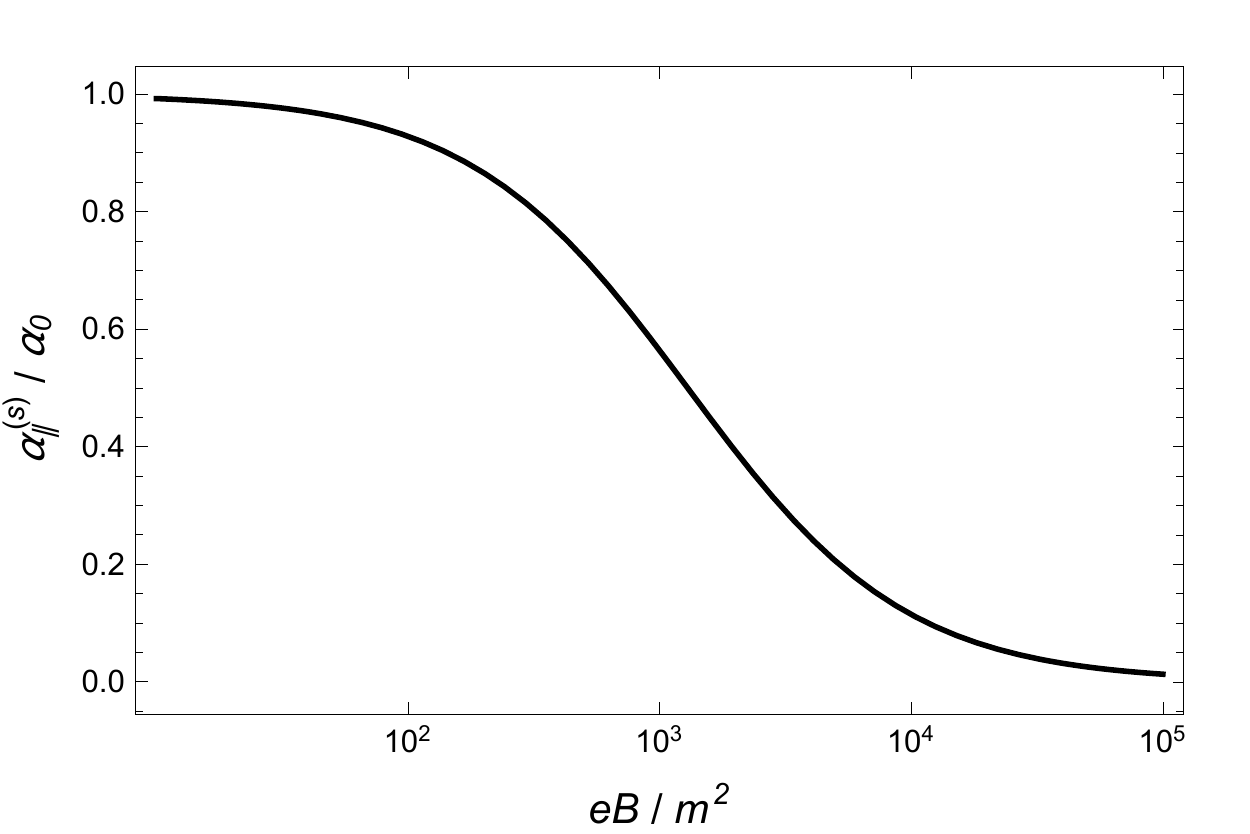} \\
    \end{center}
    \caption{Coupling constant given by Eq. (\ref{Alfa-strong-parallel})  as a function of $eB/m^2$ for $k_\perp/m=0$ and $k_{||}/m=0.5$ .}
     \label{Fig:couplingstrong}
\end{figure}

Notice that in this infrared limit,  while in the transverse direction the charge has a negligible screening, in the parallel direction, at a large distance from the charge, the effective charge decreases with the magnetic field strength (see Fig. \ref{Fig:couplingstrong}). In this case, for $|eB| \gg \frac{3 \pi}{\alpha_0} m^2$, we have that the effective charge is independent of the original coupling constant $\alpha_0$, only depending on the screening effect produced by the magnetic field, 
\begin{equation}\label{Alfa-strong-parallel-2}
\alpha^{(s)}_{QED}(k_\bot=0,B)_\| \simeq \frac{3\pi m^2}{|eB|}, \quad |eB| \gg \frac{3 \pi}{\alpha_0} m^2
\end{equation}

It is easy to check that considering $\alpha_0=1/137$ and $m=0.511$ MeV, for the electron mass, we obtain that the value of the critical field to produce this effect is $eB_c\sim 10^{16}$ G, which is a value several orders smaller than that reached in off-central heavy-ion collisions \cite{HIC}, and also smaller than the one estimated for the inner core of neutron stars \cite{EOS-Ferrer}.

Here, we should mention that our results differ from those found years ago in Refs. \cite{Chodos, Ragazzon-Alfa}, where the fine-structure constant was obtained from the Schwinger effective action in the presence of a magnetic field \cite{Proper-Time}. In Ref. \cite{Chodos}, it was found that in the strong-field approximation, $\alpha_{QED}$ moderately increases with the magnetic field, while in Ref. \cite{Ragazzon-Alfa} it was reported an anisotropic behavior, with $\alpha_{QED}$ increasing, both at strong and weak fields,  in the plane perpendicular to the magnetic field and decreasing in the direction of the field. The increase of the coupling constant in those references is found to be related with the structure $F_{\rho\lambda}F^{\rho\lambda}g_{\mu\nu}$ appearing in their photon propagator $D_{00}$,  while the decrease of the coupling was associated with a structure 
 similar to our structure $b_\mu^{(2)}b_\nu^{(2)}/(b^{(2)})^2$ in Eq. (\ref{Propagator-B}). In our case, only the second structure is present, since the first one cannot appear in the photon polarization operator because it is not transverse with respect to $k_\mu$.  This is why in the Coulomb potential energy (\ref{Coulomb-Pot-2}) we have only one coefficient (i.e. $\Pi_2$) instead of two. We point out that, as we showed in Section II, the structures of the polarization operator and photon propagator are the same in he Feynman gauge, while the Coulomb potential energy is gauge independent. Nor in Ref. \cite{Ragazzon-Alfa} , neither in \cite{Chodos}, a physical explanation of the different behavior of $\alpha_{QED}$ with the magnetic field was given. Also, because the approach in \cite{Chodos, Ragazzon-Alfa} makes use of the $k$-independent Schwinger effective action the reported fine-structure constants do not depend on the momenta.

\section{Magnetic Field dependence of the electron mass}\label{section4}

In the massive QED case, where the electron has a finite physical mass, $m_0$, the magnetic field can modify it through radiative corrections. As follows, we analyze how a weak and a strong magnetic field can affect the physical mass by taking into account its effect on the fine-structure constant. 

\subsection{Massive QED in a Weak Magnetic Field}

In the weak-field approximation, the field-dependent electron mass is known to be given as \cite{Johnson}
\begin{equation}\label{Weak-Field-Mass}
m(B)\simeq m_0\left[ 1-\frac{\alpha_0}{4\pi}\left(\frac{|eB|}{m_0^2}\right)\right]
\end{equation}

It is important to notice that this expression is only valid for magnetic-field values satisfying $1>\frac{\alpha_0}{4\pi}\left(\frac{|eB|}{m_0^2}\right)$, as has been pointed out in different contexts in Refs. \cite{Jancovici-1969, Vivian}.  For $\alpha_0$ constant, the mass decreases with the magnetic field. With the substitution of $\alpha_0$ by (\ref{Alfa-weak-1}), which is in the free-Landau-pole momentum region ($ |eB|<k^2_\bot,  k^2_3 < m_0^2$), we obtain 
\begin{equation}\label{Weak-Field-Mass}
m(B)\simeq m_0\left[ 1-\frac{\alpha_0} {4\pi[1+\frac{\alpha_0}{6\pi\textbf{k}^2}\left(\frac{|eB|}{m_0^2}\right)^2\left(\frac{k_\bot^2}{5}+\frac{k_3^2}{2}\right)-\frac{4\alpha_0}{15\pi}\left(\frac{\textbf{k}^2}{m_0^2} \right)] }\left(\frac{|eB|}{m_0^2}\right)\right]
\end{equation}

\begin{figure}
\begin{center}
  \includegraphics[width=8cm]{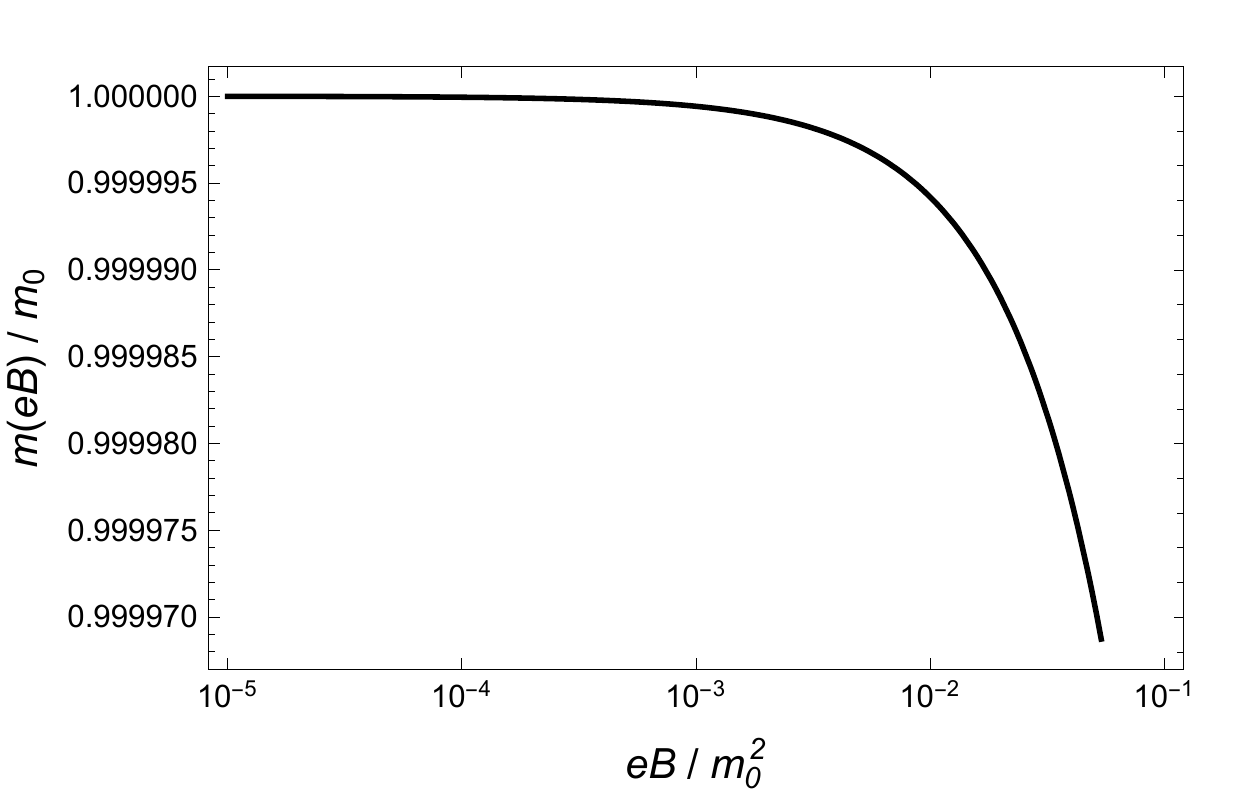} \\
    \end{center}
    \caption{Dynamical mass given by Eq. (\ref{Weak-Field-Mass})  as a function of $eB/m_0^2$ for $k_\perp/m=0.5$, $k_{||}/m=0.5$ and $\alpha_0=1/137$.}
     \label{Fig:massweakfield}
\end{figure}

In Fig. \ref{Fig:massweakfield} we plot $m(B)$ vs $|eB|/m_0^2$ for fixed values of momenta. We can notice that, increasing the magnetic field with values within the allowed region ($ |eB|<k^2_\bot,  k^2_3 < m_0^2$), $m(B)$ decreases.

\subsection{Massive QED in a Strong Magnetic Field}

 The one-loop correction in the high-magnetic-field limit ($m_0^2 \ll k_\|^2, k_\bot^2 \ll |eB|$) and under the supposition that $(\alpha_0 /4 \pi)\ln^2(|2eB|/m_0^2) \ll1$ to ensure the perturbative expansion in $\alpha_0$, was calculated in Ref. \cite{Jancovici-1969}, 
\begin{equation}\label{Perturbative-Mass-1}
m(B)\simeq m_0\left[ 1+\frac{\alpha_0}{4\pi}\ln^2\left(\frac{|2eB|}{m_0^2}\right)\right]
\end{equation}

Considering a field independent fine-structure constant, we see from (\ref{Perturbative-Mass-1}) that the mass increases with the magnetic field. Nevertheless, if we make the replacement $\alpha_0 \to \alpha^{(s)}(B)$ where $ \alpha^{(s)}(B)$ is given in Eq. (\ref{Alfa-strong-1}),  we obtain in the $m_0^2  \ll k_\|^2, k_\bot^2  \ll |eB|$ region
\begin{equation}\label{Perturbative-Mass-2}
m(k,B)\simeq m_0\left[ 1+\frac{\alpha_0} {4\pi \left(1+\frac{2\alpha_0|eB|}{\pi \textbf{k}^2} \right)}\ln^2\left(\frac{|2eB|}{m_0^2}\right)\right]
\end{equation}

While in the second region ($ k_\|^2, k_\bot^2 \ll m_0^2  \ll |eB|$) we have 
\begin{equation}\label{Perturbative-Mass-3}
m(k,B)\simeq m_0\left[ 1+\frac{\alpha_0} {4\pi \left(1+\frac{\alpha_0k_3^2|eB|}{3\pi m_0^2\textbf{k}^2} \right)}\ln^2\left(\frac{|2eB|}{m_0^2}\right)\right]
\end{equation}

\begin{figure}
\begin{center}
\begin{tabular}{ccc}
  \includegraphics[width=7cm]{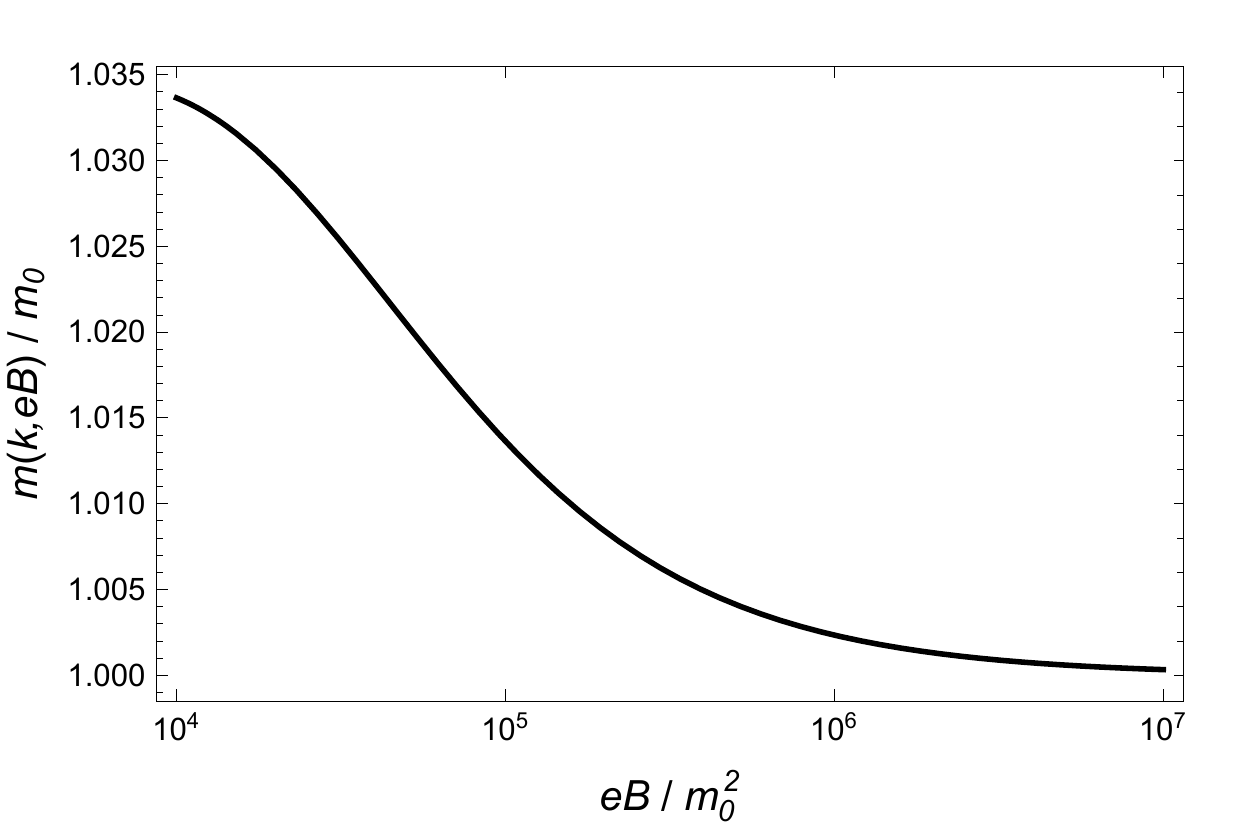} & \includegraphics[width=7cm]{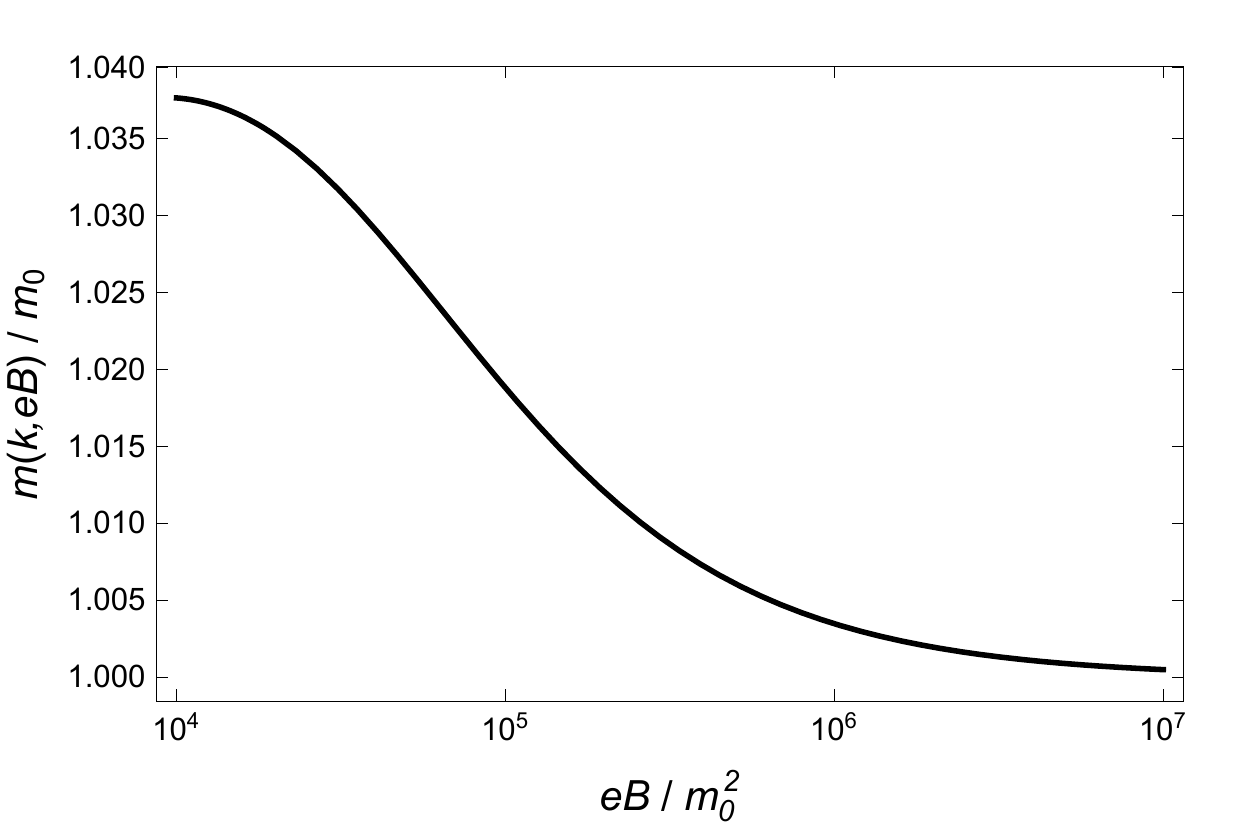}\\
(a) & (b)  \\
  \end{tabular}
    \end{center}
 \caption{Dynamical mass of Eqs.(\ref{Perturbative-Mass-2}) and (\ref{Perturbative-Mass-3})  as a function of $eB/m_0^2$ are respectively plotted in:  (a) for  $k/m_0=10$ and in  (b) for  $k/m_0=0.5$, $k_{||}/m_0=0.1$. Both with $\alpha_0=1/137$.}
    \label{Fig:massstrong}
\end{figure}

We can see from Fig. \ref{Fig:massstrong} that in both cases  $m(k,B)$ will then decrease with the magnetic field for a fixed $\textbf{k}$ value. In Fig. \ref{Fig:massstrong} we took the external momentum in the mass shell for (a) $k/m_0=10$ and  (b) $k/m_0=0.5$ and  $k_3/m_0=0.1$.

  \subsection{Massless QED and inverse magnetic catalysis}

In massless QED, the electron cannot gain a mass by radiative corrections because the chiral symmetry of the massless theory is preserved against a possible perturbative breaking.  In this case, only nonperturbative corrections can generate a dynamical mass. In the presence of a magnetic field, no important how weak it can be, if it is larger than the particle momenta, it can catalyze the chiral symmetry breaking through a phenomenon that is known in the literature as the MS$\chi$SB \cite{MC}-\cite{Paraelectricity}. The phenomenon of  MS$\chi$SB in massless QED is based on the physical idea that a magnetic field stronger than the particle momenta confines the electrons to the LLL, hence facilitating the formation of the particle/antiparticle pairs, since in the infrared region there is no gap between the LLL and the Dirac sea. As known, the condensation of this pairs endow the system quasiparticles with a dynamical mass and for those in higher Landau levels, also with an anomalous magnetic moment \cite{Ferrer-AMM-1, Ferrer-AMM-2}. 

The dimensional reduction of the LLL also contributes to the formation of the chiral condensate, since the reduction of the spatial dimension produces the strengthen of the interaction, that in this case is only carried out by longitudinal photons. Nevertheless, as we have demonstrated in this paper, a strong magnetic field produces the weakness of the coupling constant through an increase of the charge screening. If the second effect wins over the first one, then the IMC effect will be also present in this case.

 The electron dynamical mass generated through the MC$\chi$SB has been calculated by several methods. One approach \cite{Ng} is by solving the Schwinger-Dyson (SD) equation that in the presence of a magnetic field can be written as  \cite{Fradkin, Schwinger-Dyson},
\begin{equation}\label{Schwinger-Dyson Eq.} 
 \Sigma(x,y)=4\pi\alpha_0  \gamma^\mu \int G(x,x')\Gamma^\nu(x',y,y') D_{\mu \nu}(y',x)d^4x' d^4y'
\end{equation}
Here, $\Sigma (x,x')$ is the field dependent fermion self-energy operator, $D_{\mu \nu}(x-x')$ is the full photon
propagator, $G(x,x')$ is the full fermion propagator, and $\Gamma^\nu(x',p',y')$ is the full amputated vertex, which are operators depending
on the dynamically induced quantities and the magnetic field.

It can be proved \cite{Ferrer-AMM-2}, that using the Ritus Eigenfunctions \cite{Ritus:1978cj}, $E_{p}^{l}(x)$, the fermion self energy can be diagonalized in momentum space as
\begin{eqnarray}\label{P-Self-Energy}
\Sigma(p,p')= \int d^4xd^4y
\overline{E}_{p}^{l}(x)\Sigma(x,y)E_{p}^{l}(y)=(2\pi)^4\widehat{\delta}^{(4)}(p-p')\Pi(l)\widetilde{\Sigma}^l
(\overline{p}),
\end{eqnarray}
with $\widetilde{\Sigma}^l$ in the zero Landau level ($l=0$) given by the structures,
\begin{equation}\label{Structure}
 \widetilde{\Sigma}^0(\bar{p})=Z_\|^0\bar{p}_\|^\mu\gamma_\mu^\|+Z_\bot^0\bar{p}_\bot^\mu\gamma_\mu^\bot+m_{dyn}I,
\end{equation}
where the four-momentum in a magnetic field is given by  $\overline{p}^\mu=(p^{0},0, -\sqrt{2eB l},p^{3})$, $Z_\|^0$,  $Z_\bot^0$ and $m_{dyn}$ are field depending parameters, and $I$ is the unit matrix.
 
To calculate the dynamical mass it was first used the SD equation \cite{Ng} in the quenched ladder approximation, where in (\ref{Schwinger-Dyson Eq.}) it is taken the free photon propagator and bare vertex.  
We should notice however that the ladder approximation
is not gauge invariant. But as known, if the Ward-Takahashi identities are satisfied by the solution
of the SD equation in some approximation in a certain gauge, one can use
the gauge transformation law for the Green's functions \cite{WT} to rewrite the SD
equations in an arbitrary gauge. The transformation law guarantees that the
Ward-Takahashi identities are satisfied by the solutions of the SD equation in all other
gauges, although the approximation on which the SD equation is solved
may change. In the case of the ladder approximation in a magnetic field the gauge invariance of the induced chiral mass was proved through the Ward-Takahashi identities for he SD approach in the LLL limit and in the Feynman gauge in Ref. \cite{Incera}. But all this means that if we change the approximation going beyond the ladder, we will have to find of course what is the appropriate gauge where the Ward-Takahashi identities are satisfied by the solution of the SD equation. This was precisely the case in \cite{Improved-Ladder}, when considering an improved ladder approximation where the photon propagator in (\ref{Schwinger-Dyson Eq.}) was also taken full, but keeping still the bare vertex. There, it was needed a non-local gauge condition.

The solution for the dynamical mass that is obtained in the quenched ladder approximation is \cite{MC}
\begin{equation}\label{Dynamical-Mass-1}
m_{dyn}\simeq \sqrt{|eB|}\exp\left[-\frac{\pi}{2}\left(\frac{\pi}{2\alpha_0}\right)^{1/2} \right]
\end{equation}

In this approximation, as it is considered that $\alpha_0$ is constant, it is evident from (\ref{Dynamical-Mass-1}) that the dynamical mass increases with the magnetic field. Nevertheless, if we naively consider that the fine-structure constant depends on the magnetic field and make in (\ref{Dynamical-Mass-1}) the replacement $\alpha_0 \to \alpha _{QED}(B,k)$, in the strong-field approximation (\ref{Alfa-strong-1})-(\ref{Alfa-strong-2}), we obtain,
\begin{equation}\label{Dynamical-Mass-2}
m_{dyn}(k_\|, B)\simeq \sqrt{|eB|}\exp\left[-\frac{\pi}{2}\left(\frac{\pi}{2\alpha^{(s)}_{QED}(k_\bot=0,B)_\|}\right)^{1/2} \right],
\end{equation}
with
\begin{equation}\label{Alfa-strong-parallel-2}
\alpha^{(s)}_{QED}(k_\bot=0,B)_\|\simeq\frac{\alpha_0} {1+\frac{\alpha_0|eB|}{3\pi m_{dyn}^2}},   \quad k^2_\|<m_{dyn}^2 < |eB|,
\end{equation}

\begin{equation}\label{Alfa-strong-12}
\alpha^{(s)}_{QED}(k_\bot=0,B)_\|\simeq\frac{\alpha_0} {1+\frac{2\alpha_0|eB|}{\pi k^2_\|}}, \quad m_{dyn}^2<k^2_\| < |eB|,
\end{equation}
In (\ref{Dynamical-Mass-2})-(\ref{Alfa-strong-12}) we took into account that in the LLL the fermions only interchange longitudinal momentum with the photon fields (i.e. $k_\bot^2=0$).

To find how the dynamical-mass varies with the field, it is necessary to solve the self-consistent system of coupled equations (\ref{Dynamical-Mass-2}) and (\ref{Alfa-strong-parallel-2}) for the region, $ k^2_\|<m_{dyn}^2 < |eB|$, and (\ref{Dynamical-Mass-2}) and (\ref{Alfa-strong-12}) for the region, $ m_{dyn}^2<k^2_\| < |eB|$. Substituting (\ref{Alfa-strong-parallel-2}) into (\ref{Dynamical-Mass-2}), it is easy to find that there is no solution for the dynamical mass as a function of the magnetic field. While if $\alpha^{(s)}_{QED}(k_\bot=0,B)_\|=\alpha_0$, it is found that the mass is almost independent on the momentum up to $|k|\leqslant m_{dyn}(k=0)$, from where it begins to rapidly decrease \cite{Emilio}.

For the second parameter region, $m_{dyn}^2<k^2_\| < |eB|$, let us substitute (\ref{Alfa-strong-12}) into  (\ref{Dynamical-Mass-2}) and normalize the resultant equation with respect to the the electron critical field $B_c^{(e)}$,
\begin{equation}\label{Dynamical-Mass-22}
\frac{m_{dyn}(k_\|, B)}{\sqrt{|eB_c^{(e)}|}}= \frac{\sqrt{|eB|}}{\sqrt{|eB_c^{(e)}|}} \exp\left[-\frac{\pi}{2}\left(\frac{\pi}{2\alpha_0}+ \frac{\frac{|eB|}{|eB_c^{(e)}| }}{\frac{k^2_\|} {|eB_c^{(e)}| }}\right)^{1/2} \right],
\end{equation}

\begin{figure}
\begin{center}
\begin{tabular}{ccc}
  \includegraphics[width=7cm]{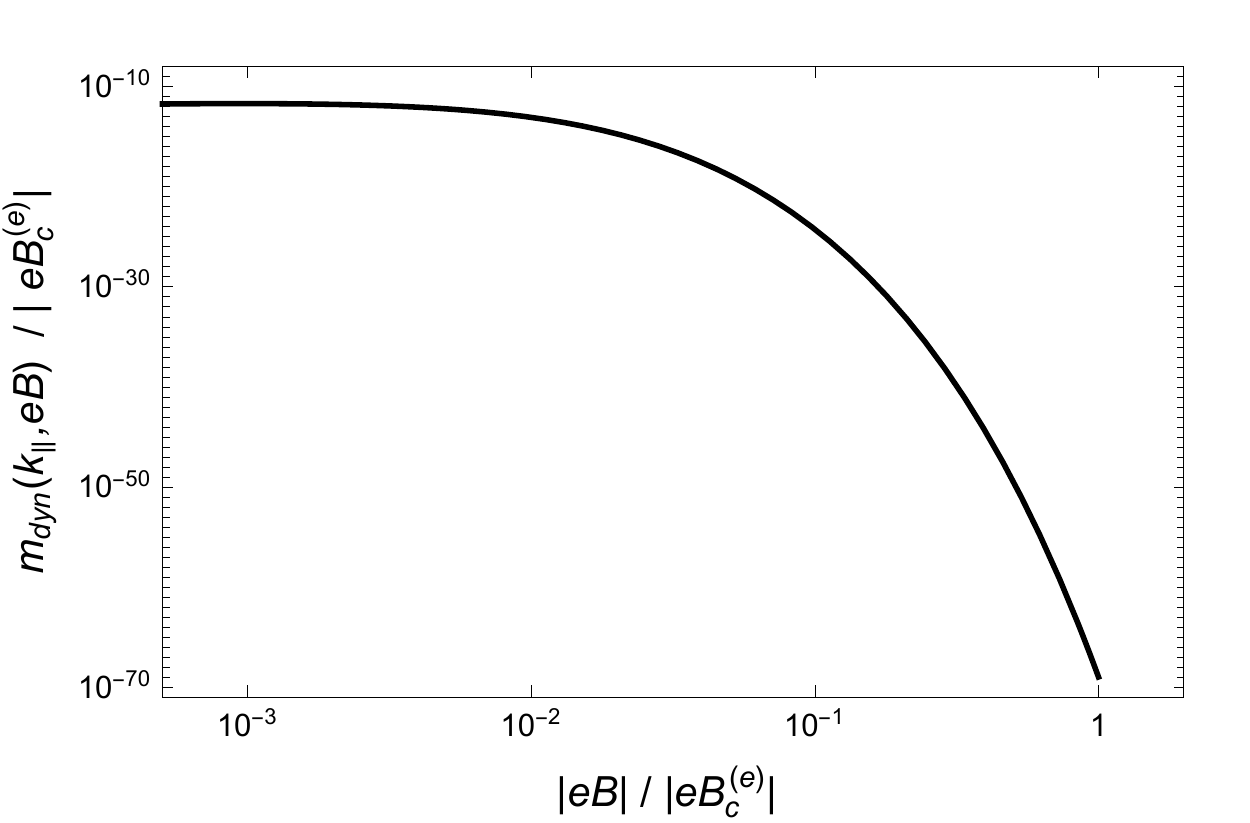} & \includegraphics[width=7cm]{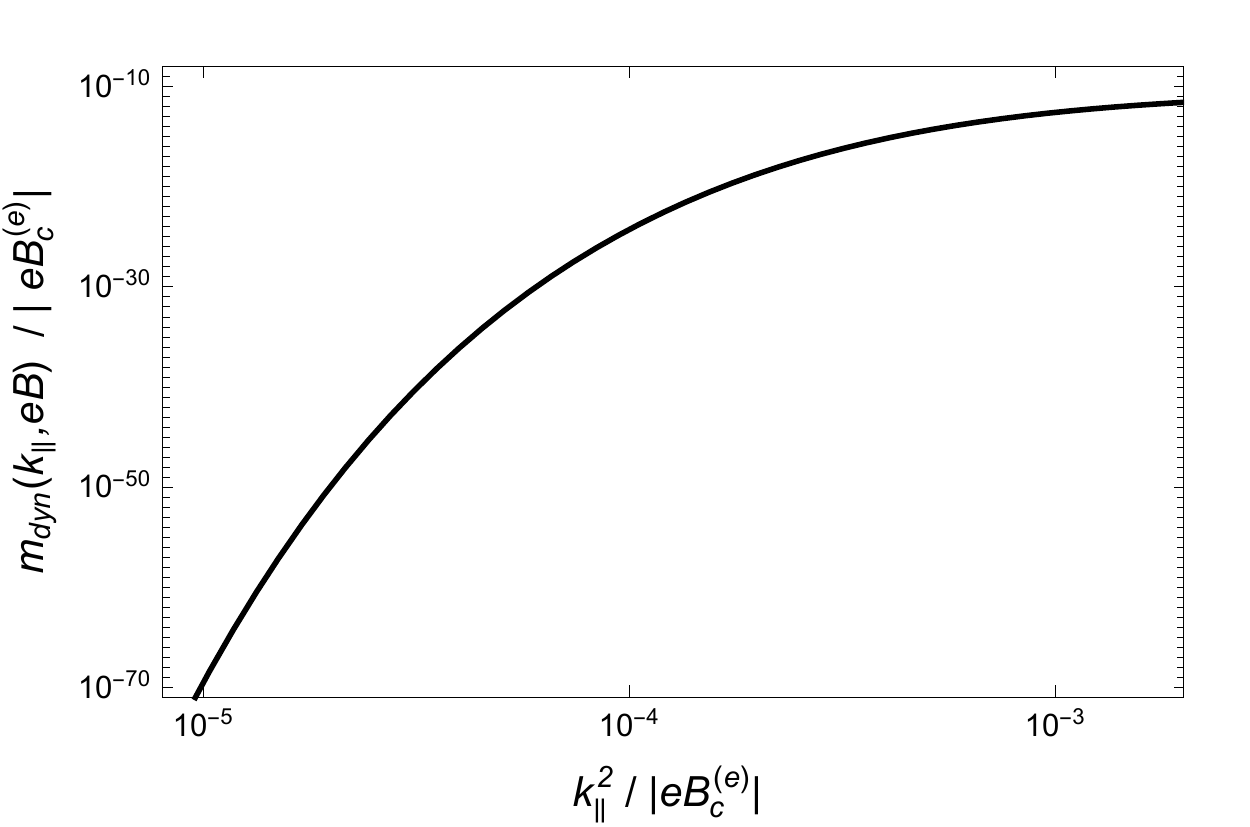}\\
(a) & (b)  \\
  \end{tabular}
    \end{center}
    \caption{Dynamical mass behavior of Eq.(\ref{Dynamical-Mass-22}): (a) as a function of $eB/|eB^{(e)}_c|$ with $k_{||}^2/|eB_c^{(e)}|=10^{-4}$, and (b) as a function of $k_{||}^2/|eB^{(e)}_c|$ with $eB/|eB_c^{(e)}|=10^{-1}$ .}
     \label{Fig:couplingweaklimit-paraperp}
\end{figure}

In Fig. \ref{Fig:couplingweaklimit-paraperp}a we plot how the dynamical mass changes with the field at a fixed value of the longitudinal momentum, and in Fig. \ref{Fig:couplingweaklimit-paraperp}b, how the dynamical mass changes with the longitudinal momentum at a fixed value of the applied field. Notice that the behavior of the dynamical mass in this case is totally opposite to that obtained for a constant coupling constant. Moreover, taking into account that the critical temperature to regain the chiral symmetry in the system, $T_c$, is proportional to the value of the electron dynamical mass at zero temperature \cite{Gusynin}, we have that the decrease of $m_{dyn}$ with the magnetic field will produce a decrease of $T_c$, which is the typical behavior of the IMC phenomenon in QCD \cite{Inverse-MC}. In the following section we give a physical explanation for all these peculiar behaviors and discuss the limitations of these results.

\section{Physical discussion and Summary}\label{section5}

As it is generally accepted, to understand the particle dynamics in quantum field theory, it is necessary to know how the different physical parameters, as particle momentum, temperature, electromagnetic fields, etc., affect the coupling constants. In this paper we are in particular interested in the effect of an applied uniform and constant magnetic field in the fine-structure constant and consequently into the electron mass in two cases: in normal QED, where there is a different from zero electron mass that can be affected by a magnetic field through radiative corrections, and in massless QED, where the electron mass can be dynamical generated by the so called MC$\chi$SB phenomenon \cite{MC}-\cite{Paraelectricity}. In the second case, we are considering the dynamical mass calculated in the rainbow approximation, as a first look to the IMC phenomenon in QED.

We have found that contrary to other results previously reported in the literature \cite{Chodos, Ragazzon-Alfa}, the fine-structure constant in massive QED decreases with the magnetic field in the weak-field, as well as in the strong-field limits. To understand why this is the physical result to be expected, let's start by considering the zero-field situation. In this case, the fine-structure constant decreases toward larger distances. The decrease is produced by the vacuum polarization effect  due to the electron-positron pairs that can be continuously created from the vacuum in faith of the Heisenberg uncertainty principle. The pairs produce a screening effect that increases with distance as the cloud of this virtual particles increases. Now, when we apply a magnetic field, the virtual particles redistribute in Landau levels, each of which has a state degeneracy. As known, increasing the field, the density of states of each Landau level to be occupied  by the virtual pairs increases, so the screening corresponding to a fixed distance also increases, producing the decrease of the coupling constant at that distance.

In making the physical analysis of the investigated scenarios, we should take into account the three main scales that are involved: the magnetic length, $l_M\sim 1/\sqrt{|eB|}$, which is associated with the radius of the LLL; the Compton wavelength, $l_C\sim1/m$, which is associated with the quantum field theory region where the particle-antiparticle pairs are created; and the observation length, $l_O\sim 1/|\textbf{k}|$, from where we want to define the charge effective value.

In the strong-field limit, the coupling constant becomes significantly anisotropic. In the direction transverse to the field, the coupling constant does not vary with the field neither the momentum. In this case the virtual cloud is confined to the LLL without any freedom to move in the transverse direction by virtue of jumping between Landau levels. For the coupling in the longitudinal direction, we found in the infrared region, $l_M < l_C <l_O$, that when the field is strong enough $|eB|>\frac{3\pi}{\alpha_0}m^2$, the effective coupling constant becomes independent of the original charge and of the observation distance along the field direction. In this case, there is a strong screening that only depends on the magnetic field and particle mass. The decrease of the effective coupling with the field can be seen here as a consequence of the fact that since the magnetic length is smaller than the Compton wavelength, the virtual pairs are confined to the LLL. Then, by increasing the field, the magnetic length decreases, and at the same time, the state degeneracy increases.  Hence, for any observation distance along the magnetic field direction, which is larger than the Compton wavelength, the screening is the same and the net charge that is seen is almost zero due to the strong screening. 

As it is known, in perturbative QED at zero magnetic field there exists a charge singularity at short distance that is called the Landau pole \cite{Landau-pole}. This singularity hinders perturbative QED at very short distances. 
Now, once a magnetic field is applied, we have shown that even in the weak-field limit (\ref{Alfa-weak-2}), the effect of the magnetic field is to counteract the singularity, although the magnetic field strength in the weak-field case is not enough to remove it. Nevertheless, at strong field, when the magnetic-field nonperturbative contribution is taken  into account in the one-loop calculation of the fine-structure constant, the pole disappears, as can be seen from (\ref{Alfa-strong-1}) at $l_O<l_C$. In this case, when all the LLL states are populated at $l_M\ll l_O$, the effective coupling is completely screened.

In the weak-field limit, $l_M > l_C$, it is regained the zero-field limit of the coupling constant as the leading contribution (\ref{Alfa-weak-1})-(\ref{Alfa-weak-2}), while the magnetic field produces a weak screening effect.  In this case, the leading contribution is coming from the virtual cloud that is smeared through out the whole space since the separation between Landau levels is so tiny that resembles a continue distribution. 

When calculating the radiative effect of a magnetic field on the electron mass, the field effect on the fine-structure constant should be considered. In the case of massive QED, the role of a strong magnetic field is important and the screening effect in the coupling becomes significant. Hence, the net effect of the applied magnetic field gives a mass decrease with the field. In the weak-field limit, the screening effect persists, but in a lower degree.

For massless QED the Compton wavelength becomes a dynamical parameter, $l_C\sim1/m_{dyn}$, that depends on the magnetic field and becomes very large for a small $m_{dyn}$. Thus, the magnetic length is always smaller than the Compton wavelength ($l_M<l_C$), and the states in the Landau levels available for the virtual cloud should be affected by the magnetic field, and consequently affecting the screening. In the used approximation, we have found that this screening effect results more important than the strengthen of the interaction due to the spatial dimensional reduction in the LLL and this is why the dynamical mass decreases with the field as shown in Fig. \ref{Fig:couplingweaklimit-paraperp}. 
Consequently, the critical temperature needed to regain the chiral symmetry decreases with the field, since it is proportional to the dynamical mass. This is the distinctive signal of IMC in this system.
On the other hand, the dynamical mass increases with the longitudinal momentum because the coupling constant increases with the momentum as it is typical for a theory without asymptotic freedom. However, the results we are presenting here for the massless case only serve to give a signal of the importance of the role the running coupling can play in the field dependence of the dynamically induced mass. This role is missing in the ladder and improved ladder approximations considered up to now, because there, it has been considered a bare vertex. But in order to have a consistent treatment for this problem we need to go beyond those approximations and considerer in the SD equation 
(\ref{Schwinger-Dyson Eq.}) a full vertex together with the full propagators that will all depend on the magnetic field and dynamical mass. For this approximation to be reliable the appropriate gauge condition that makes the SD solution to satisfy the Ward-Takahashi identities should be found, what is not a trivial task.

\acknowledgments
EJF work was supported in part by NSF grant PHY-1714183. AS work was supported in part by DGAPA-UNAM  grant  PAPIIT-IN118219.

\appendix
\section{$\Pi_2$ coefficient of the one-loop photon propagator at $B\neq0$}\label{appendixA-1}

\subsection{Strong-Field Limit}\label{strong-app}

Our goal now is to find the one-loop polarization operator in the strong-field limit. We will use the Ritus's method \cite{Ritus:1978cj}, where the Landau levels appear in an explicit way. Then, for a weak coupled theory, the strong-field limit prescription results in keeping only the LLL contribution, since at a strong field the particles will be confined into their lowest energy state (i.e. they will not have enough energy to jump across the energy gap separating the Landau leves that is proportional to $\sqrt{|eB|}$). We will show, that working in this form, it is obtained the same result that was found in \cite{Batalin} by using the Schwinger proper-time method where the sum in all Landau levels was considered.

The photon polarization operator in the one-loop approximation reads
\bea
   \Pi_{\mu\nu}(x,y)
        &=&-4\pi\alpha\ Tr\left[
           \gamma_\mu G(x,y)\gamma_\nu G(y,x)
            \right]\,
\label{oppol1}
\eea
where  $G(x,y)$ is the electron propagator, and $\gamma^\mu$ the Dirac's gamma matrices.

Taking into account that in the momentum space the electron propagator has the form
\bea
    G(x,x')=\sumint\frac{d^4p}{(2\pi)^4}
           \mathbb{E}_p(x)\Pi(l)
           \frac{1}{\slsh{\overline{p}}-m}
           \overline{\mathbb{E}}_p(x')\,
\label{oppol2}
\eea
where, in the Landau gauge $A^\mu=(0,0,Bx^1,0)$, the 4-momentum is $\overline{p}^\mu=(p^0.0,-\sqrt{2eBl},p^3)$ and $\mathbb{E}_p$ are the Ritus's eigenfunctions matrices given by
\bea
  \mathbb{E}_p&=&\sum_{\sigma=\pm1}E_{p\sigma}(x)\Delta(\sigma),
\label{ad2}
\eea
with
\bea
   E_{p\sigma}(x)=N_n e^{i(p_0x^0+p_2x^2+p_3x^3)}D_n(\rho)\, ,
\label{ad4}
\eea
Here, $D_n(\rho)$  denotes the parabolic cylinder functions with argument
$\rho=\sqrt{2|eB|}(x^1-p^2/eB)$, normalization factor $N_n=(4\pi
eB)^{1/4}/\sqrt{n!}$,  and  positive integer index
\bea
   n=n(l,\sigma)\equiv l+\frac{\sigma+1}{2}\, .
\label{ad5}
\eea
The spin projector are defined as
\bea
   \Delta(\sigma)\equiv\frac{I+ i\sigma \gamma^1\gamma^2}{2}\, ,
\label{ad3}
\eea
 Also, the factor $\Pi(l)\equiv\Delta(+)+\Delta(-)(1-\delta_{0l})\,$ and  the notation
\bea
    \sumint\frac{d^4p}{(2\pi)^4}\equiv
    \sum_l\int\frac{dp^0dp^2dp^3}{(2\pi)^4}\,
\label{oppol3}
\eea
wwere introduced.

Then, \Eq{oppol1} in  the momentum space reads
\bea
    &&\hspace{-0.3cm}
      \Pi_{\mu\nu}(k)=-2\alpha eB
       e^{-\hat{k}^2_\perp}
      \sum_{l,l'}\int \frac{d^2p}{(2\pi)^2}
      \sum_{\{\sigma\}}
      \frac{e^{i(n-n'+\overline{n}'-\overline{n})\phi}}
           {\sqrt{n!n'!\overline{n}!\overline{n}'!}}
       \frac{J_{nn'}(\hat{k}_\perp) J_{\overline{n}'\overline{n}}(\hat{k}_\perp)}{[\overline{p}^2-m^2][(\overline{p-k})^2-m^2]}
\nonumber \\
       &&\hspace{3.5cm}\times
       Tr\left[
         \Delta(\overline{\sigma}')\gamma_\mu\Delta(\overline{\sigma})
         \Pi(l)(\overline{p}+m)
        \Delta(\sigma)\gamma_\nu\Delta(\sigma')
         \Pi({l'})(\slsh{\overline{p}}-\slsh{\overline{k}}+m)
       \right]
\label{oppol9}
\eea
where $\{\sigma\}$ means sum over $\sigma$, $\sigma'$,
$\overline{\sigma}$ and $\overline{\sigma}'$, and $\overline{p-k}\ ^\mu \equiv(p^{0}-k^{0},0,
-\sqrt{2eBl'},p^{3}-k^{3})$.

In the derivation of Eq.(\ref{oppol9}), we used the identities~\cite{leeleungng}
\bea
  &&\hspace{-1.5cm}\int dy \ e^{-i k'\cdot y}
    \overline{\mathbb{E}}_p(y)\gamma_\nu
    \mathbb{E}_{p'}(y)
    \nonumber \\
   &&=(2\pi)^4 \hat{\delta}^{(3)}(p'+k'-p)
   e^{-\frac{\hat{k}'^2}{2}}
   e^{-i\frac{k_1'(p_2'+p_2)}{2eB}}
    \sum_{\sigma, \sigma'}\frac{J_{nn'}(\hat{k}'_\perp)e^{i(n-n')\phi}}{\sqrt{n!n'!}}
    \Delta(\sigma)\gamma_\nu\Delta(\sigma')
\label{oppol6}
\eea
and
\bea
   &&\hspace{-1.5cm}\int dx\ e^{i k\cdot x}
     \overline{\mathbb{E}}_{p'}(x)\gamma_\mu
     \mathbb{E}_{p}(x)
\nonumber \\
    &&= (2\pi)^4 \hat{\delta}^{(3)}(p'+k-p)
    e^{-\frac{\hat{k}^2}{2}}
    e^{i\frac{k_1(p_2'+p_2)}{2eB}}
    \sum_{\overline{\sigma},\overline{\sigma}'}
    \frac{J_{\overline{n}'\overline{n}}(\hat{k}_\perp)e^{i(\overline{n}'-\overline{n})\phi}}
    {\sqrt{\overline{n}!\overline{n}'!}}
    \Delta(\overline{\sigma}')\gamma_\mu\Delta(\overline{\sigma})\,,
\label{oppol7}
\eea
where $n=n(l,\sigma)$, $n'=n(l',\sigma')$,
$\overline{n}=n(l,\overline{\sigma})$ and
$\overline{n}'=n(l',\overline{\sigma}')$ with $n$ given by \Eq{ad5},
and
\bea
    J_{nn'}(\hat{k}_\perp)
     \equiv\sum_{m=0}^{min(n,n')}
           \frac{n!n'!\,\,|i\hat{k}_\perp|^{n+n'-2m}}{m!(n-m)!(n'-m)!}\,
\label{jeqn}
\eea
with $\hat{k}_\perp\equiv k_\perp/2eB$.

In the low energy region or strong field limit, $\hat{q}\ll 1$,  only those terms with smallest power in
$\hat{q}_\perp$ in $J_{nn'}$ and $J_{\overline{n}'\overline{n}}$ contribute
to the polarization operator. Then, in the leading approximation
\bea
   J_{nn'}(\hat{q}'_\perp)
      \rightarrow n!\delta_{nn'}
\hspace{0.5cm}\mbox{and} \hspace{0.5cm}
   J_{\overline{n}'\overline{n}} \rightarrow \overline{n}!
    \delta_{\overline{n}'\overline{n}}\,,
\label{oppol11}
\eea
and taking $l=l'=0$, the photon polarization operator in the LLL approximation has the form
\bea
     \Pi^{||}_{\mu\nu}(k)&=&-2\alpha eBe^{-\hat{k}^2_\perp}
       \int \frac{dp^0dp^3}{(2\pi)^2}
       \frac{
       Tr\left[
           \gamma^{||}_\mu
           \Delta(+)(\slsh{\overline{p}}^{||}+m)
           \gamma^{||}_\nu
           \Delta(+)(\slsh{\overline{p}}^{||}-\slsh{\overline{k}}^{||}+m)
       \right]}{[p^2_{||}-m^2][(p-k)_{||}^2-m^2]}
\label{oppol63}
\eea
where $\gamma^{\mu}_{||}=(\gamma^0,0,0,\gamma^3)$ and $\overline{p}^\mu_{||}=(p^0,0,0,p^3)$.

The trace over Dirac gamma matrices is  straightforward and we get
\bea
   &&\hspace{-3cm}Tr\left[
           \gamma^{||}_\mu
           \Delta(+)(\slsh{\overline{p}}^{||}+m)
           \gamma^{||}_\nu
           \Delta(+)(\slsh{\overline{p}}^{||}-\slsh{\overline{k}}^{||}+m)
       \right]      
       \nonumber \\
       &&=2
        \left(
          \overline{p}_\mu^{||}(\overline{p}-\overline{k})_\nu^{||}
         +\overline{p}_\nu^{||}(\overline{p}-\overline{k})_\mu^{||}
         -g_{\mu\nu}(\overline{p}^{||}\cdot(\overline{p}-\overline{k})^{||}-m^2)
        \right).
\label{oppol26}
\eea
Replacing \Eq{oppol26} in \Eq{oppol63}, we obtain
\bea
     \Pi^{||}_{\mu\nu}(k)&=&
      -2\alpha eB
       e^{-\hat{k}^2_\perp}
       \int \frac{d^2p_{||}}{(2\pi)^2}
       \frac{ 2\left[\overline{p}_\mu^{||}(\overline{p}-\overline{k})_\nu^{||}
         +\overline{p}_\nu^{||}(\overline{p}-\overline{k})_\mu^{||}
         -g_{\mu\nu}(\overline{p}^{||}\cdot(\overline{p}-\overline{k})^{||}-m^2)\right]
         }{[p_{||}^2-{m}^2][(p-k)_{||}^2-m^2]}.
\nonumber \\
\label{oppol34}
\eea
The integration over momenta can be easily carried out by using Feynman
parametrization, it is
\bea
     \Pi^{||}_{\mu\nu}(k)&=&
      -2\alpha eB
       e^{-\hat{k}^2_\perp}
       \int _0^1 dx\int \frac{d^2l_{||}}{(2\pi)^2}
       \frac{2}{[l_{||}^2-\mathcal{M}]^2}
     \left[
           2l_\mu^{||}l_\nu^{||}
           -2x(1-x)k_\nu^{||}k_\mu^{||}
           -g^{||}_{\mu\nu}(l_{||}^2+\mathcal{M}-2{m}^2)
            \right]
\nonumber\\
\label{feynpara}
\eea
where $l_{||}^\mu\equiv p_{||}^\mu-(1-x)k_{||}^\mu$ and
$\mathcal{M}\equiv-x(1-x)k_{||}^2+{m}^2$. 

By using dimensional regularization, we remove the logarithmic divergent part, and we  obtain
\bea
  \Pi^{||}_{\mu\nu}(k)&=&
     i\left(g^{||}_{\mu\nu}-\frac{k_\mu^{||}k_\nu^{||}}{k_{||}^2}
       \right)\Pi(k)
\label{oppol68}
\eea
with
\bea
   \Pi(k)&\equiv&-\frac{2\alpha eB}{\pi} e^{-\hat{k}^2_\perp}
    \int_0^1\ dx  \frac{x(1-x)k_{||}^2}{-x(1-x)k_{||}^2+m^2}
\nonumber \\
       &=&\frac{2\alpha eB}{\pi} 
             e^{-\hat{k}^2_\perp}
             \left[1+\frac{2m^2}{\sqrt{k_{||}^2(k_{||}^2-4m^2)}}
                         \ln\left(\frac{\sqrt{4m^2-k_{||}^2}-\sqrt{-k_{||}^2}}{\sqrt{4m^2-k_{||}^2}+\sqrt{-k_{||}^2}}\right)
             \right]
\label{intlog}
\eea
where the $i$ factor in (\ref{oppol68}) comes from the integration over the momentum in
the Minkowski space~\cite{peskin}.

Since the tensor structure in Eq.(\ref{oppol68}) can be written in terms of  $b^{(2)}_\mu b^{(2)}_\nu$  (see Eq. (\ref{Egin-Vectors})) as
\begin{equation}
   g_{\mu\nu}^{||}-\frac{k^{||}_{\mu}k^{||}_{\nu}}{k_{||}^2}=\frac{b^{(2)}_\mu b^{(2)}_\nu}{{(b^{(2)})}^2}    
\end{equation}
with $g^{||}_{\mu\nu}=diag(1,0,0,-1)$, then, the coefficient $\Pi(k)$ in Eq.(\ref{oppol68}) is no other than $\Pi_2(k,B)$ in Eq.(\ref{Polarization-Opr}). 

In the region $k_{||}^2\ll m^2 \ll |eB|$, $\Pi_2(k,B)$ has the asymptotic behavior
\bea
    \Pi_2(k,B)= \frac{\alpha k_3^2 |eB|}{3\pi m^2}e^{-\hat{k}^2_\perp}
\eea
and  for the region $ m^2\ll k_{||}^2\ll  |eB|$, $\Pi_2(k,B)$ behaves as
\bea
    \Pi_2(k,B)= \frac{2 \alpha|eB|}{\pi}e^{-\hat{k}^2_\perp}.
\eea

These results coincide with those reported in Refs. \cite{Batalin, Shabad-Trudy} where it was used the Schwinger proper-time approach.

\subsection{Weak-Field Limit}\label{weak-app}

The general expression of the $\Pi_2(0, \textbf{k},B)$ coefficient was obtained from the one-loop polarization operator in the proper-time approach in Ref. \cite{Shabad-JETP}.  Considering a uniform and constant magnetic field along the $x_3$ direction it was given by
\begin{equation}\label{Coefficien-WA}
 \Pi_2(k,B)=-\frac{1}{2}(k^2_\|\Sigma_2+k^2_\bot\Sigma_1)
 \end{equation}
with $k^2_\|=k_3^2-k_0^2$, and $k^2_\bot=k_1^2+k_2^2$. Notice in (\ref{Coefficien-WA}) the anisotropy introduced by the uniform magnetic field between the longitudinal and transverse momentum components. 

In (\ref{Coefficien-WA}), the following notation was introduced,
\begin{equation}\label{Coefficient-Sigma}
\Sigma_i=\Sigma_i^{(1)}+\Sigma_i^{(2)},
 \end{equation}
 
 \begin{equation}\label{Coefficient-Sigma-1}
\Sigma_i^{(1)}=\frac{2\alpha}{\pi}\int_0^\infty e^ {-B_{cr}t/B}\left(\frac{g_i(t)}{\sinh(t)}-\frac{1}{3t}\right)dt
\end{equation}

 \begin{equation}\label{Coefficient-Sigma-2}
\Sigma_i^{(2)}=\frac{2\alpha}{\pi}\int_0^\infty e^ {-B_{cr}t/B} dt \int_{-1}^1d\eta  \frac{\sigma_i(t,\eta)}{\sinh(t)} \left[exp\left(-k_\bot^2\frac{M(t,\eta)}{eB}-k_\|^2\frac{1-\eta^2}{4eB}t\right)-1\right]
\end{equation}
where $B_{cr}=m^2/e$ and
\begin{equation}\label{Coefficients-g}
g_1(t)=\frac{1}{4t\sinh(t)} \left(\frac{\sinh(2t)}{t}-2 \right), \quad g_2(t)=\frac{\cosh(t)}{3},   
\end{equation}

\begin{equation}\label{Coefficients-sigma}
\sigma_1(t,\eta)=\frac{1-\eta}{2} \frac{\sinh[(1+\eta)t]}{2\sinh(t)}, \quad \sigma_2(t,\eta)=\frac{1-\eta^2}{4}\cosh(t), 
\end{equation}

\begin{equation}\label{Coefficients-M}
M(t,\eta)=\frac{\cosh(t)-\cosh(t\eta)}{2\sinh(t)} 
\end{equation}

Our goal now is to get the weak-field limit of (\ref{Coefficien-WA}). It is easy to check that for $B_{cr}/B\gg 1$ all the integrands in (\ref{Coefficient-Sigma-1})-(\ref{Coefficient-Sigma-2}) are suppressed at high $t$ ($t > B/B_{cr}$) due to the exponential damping $e^{-B_{cr}t/B}$. Thus, the leading contribution at weak-field is found in the region $0\leqslant t \leqslant B/B_{cr} \ll 1$. In this small $t$ region the asymptotic behavior of the parameters are given by

\begin{equation}\label{Coefficients-g2}
g_1(t\sim 0)=\frac{1}{3}+\frac{t^2}{15}, \quad g_2(t\sim 0)=\frac{1}{3}+\frac{t^2}{6},   
\end{equation}

\begin{equation}\label{Coefficients-sigma2}
\frac{\sigma_i(t\sim 0,\eta)}{\sinh(t\sim 0)}=\frac{1-\eta^2}{4t}, \quad i=1,2
\end{equation}

\begin{equation}\label{Coefficients-M2}
M(t\sim 0,\eta)=\frac{1-\eta^2}{4}t
\end{equation}

Plugging (\ref{Coefficients-g2})-(\ref{Coefficients-M2}) into (\ref{Coefficient-Sigma-1}) and (\ref{Coefficient-Sigma-2}) we obtain
\begin{equation}\label{Coefficient-Sigma-11-weak}
\Sigma_1^{(1)}\simeq \frac{2\alpha}{\pi} \int_0^{B/B_{cr}} \left[\frac{1}{t}\left(\frac{1}{3}+\frac{t^2}{15}\right)-\frac{1}{3t}\right]dt= \frac{\alpha}{15\pi}\left(B/B_{cr}\right)^2, 
\end{equation}

\begin{equation}\label{Coefficient-Sigma-21-weak}
 \Sigma_2^{(1)}\simeq \frac{2\alpha}{\pi} \int_0^{B/B_{cr}}\left(\frac{1}{3t}+\frac{t}{6}-\frac{1}{3t}\right)dt=\frac{\alpha}{6\pi}\left(B/B_{cr}\right)^2,
\end{equation}

\begin{eqnarray}\label{Coefficient-Sigma-22-weak}
\Sigma_1^{(2)}=\Sigma_2^{(2)}\simeq\frac{2\alpha}{\pi} \int_0^{B/B_{cr}} dt\int_{-1}^1d\eta \frac{1-\eta^2}{4t}\left[e^{-k_\|^2(\frac{1-\eta^2}{4eB})t-k^2_\bot (\frac{1-\eta^2}{4eB})t}-1\right]
\nonumber
\\
=\frac{\alpha}{2\pi}\int_{-1}^1d\eta (1-\eta^2)\int_0^{x_0} dx \frac{e^{-x}-1}{x}=\frac{\alpha}{2\pi}\int_{-1}^1d\eta(1-\eta^2)\left[Ei(-x_0)-C-ln(x_0)\right]
\end{eqnarray}
where $x_0=\frac{1-\eta^2}{4}\frac{k^2}{m^2}$, $Ei(x)$ is the exponential-integral function and $C$ the Euler constant. In (\ref{Coefficient-Sigma-22-weak}) we used the formula \cite{Gradshteyn}

 \begin{equation}\label{Gradshteyn}
\int_0^{x_0} dx \frac{e^{-x}-1}{x}=Ei(-x_0)-C-ln(x_0)
\end{equation}

To integrate (\ref{Coefficient-Sigma-22-weak}) in $\eta$, we have to consider two cases:

a) \textbf{$k/m\ll1$}

In this case we use the formula \cite{Gradshteyn}
\begin{equation}\label{Gradshteyn-2}
Ei(-x_0)=C+ln(x_0)+\sum_{n=1}^\infty(-)^n\frac{x_0^n}{n n!}
\end{equation}
to write (\ref{Coefficient-Sigma-22-weak}) as
\begin{equation}\label{Coefficient-Sigma-22-weak-2}
\Sigma_1^{(2)}=\Sigma_2^{(2)}\simeq \frac{\alpha}{2\pi}\int_{-1}^1d\eta \sum_{n=1}^\infty \frac{(-)^n}{n n!} \left(\frac{1-\eta^2}{4}\right)^{n+1}\left(\frac{k}{m}\right)^{2n}
\end{equation}

Hence, the leading contribution is given by
\begin{equation}\label{Coefficient-Sigma-22-weak-3}
\Sigma_1^{(2)}=\Sigma_2^{(2)}\simeq -\frac{\alpha}{2\pi}\frac{k^2}{m^2}\int_{-1}^1d\eta \left(1-\eta^2\right)^2=-\frac{8\alpha}{15\pi}\frac{k^2}{m^2}
\end{equation}

Substituting with (\ref{Coefficient-Sigma-11-weak}), (\ref{Coefficient-Sigma-21-weak}) and (\ref{Coefficient-Sigma-22-weak-3}) into (\ref{Coefficient-Sigma}) we obtain
\begin{equation}\label{Sigma-1}
\Sigma_1=\Sigma_1^{(1)}+\Sigma_1^{(2)}\simeq \frac{\alpha}{15\pi}\left(\frac{eB}{m^2}\right)^2-\frac{8\alpha}{15\pi}\frac{k^2}{m^2}
\end{equation}

\begin{equation}\label{Sigma-2}
\Sigma_2=\Sigma_2^{(1)}+\Sigma_2^{(2)}\simeq \frac{\alpha}{6\pi}\left(\frac{eB}{m^2}\right)^2-\frac{8\alpha}{15\pi}\frac{k^2}{m^2}
\end{equation}

Now, plugging in (\ref{Coefficien-WA}) the results (\ref{Sigma-1}) and (\ref{Sigma-2}) we have
\begin{equation}\label{Coefficien-WA-1}
 \Pi_2^{(w)}(eB<k_\|^2, k_\bot^2<m^2)=-\frac{1}{2}\left[\frac{\alpha}{3\pi}\left(\frac{eB}{m^2}\right)^2\left(\frac{k_\bot^2}{5}+\frac{k_\|^2}{2}\right)-\frac{8\alpha}{15\pi}\frac{k^4}{m^2}\right]
 \end{equation}

b) $k/m\gg1$

Here it is convenient to take in (\ref{Coefficient-Sigma-22-weak}) the asymptotic expansion of the exponential-integral function \cite{Magnus}
\begin{equation}\label{Sigma-2-1}
Ei(-x_0)=-\frac{e^{-x_0}}{x_0}\left[\sum_{n=0}^N\frac{n!}{(-x_0)^n} +O(|x_0|^{-N-1})\right], \quad |x_0|\gg 1
\end{equation}

Then, the leading term in (\ref{Coefficient-Sigma-22-weak}) is given by the logarithm
\begin{equation}\label{Sigma-2-2}
\Sigma_1^{(2)}=\Sigma_2^{(2)}\simeq-\frac{2\alpha}{\pi}\int_{-1}^1d\eta \frac{1-\eta^2}{4}\ln \left(\frac{1-\eta^2}{4}\frac{k^2}{m^2}\right)\simeq -\frac{4\alpha}{3\pi}\ln\left(\frac{k}{m}\right)
\end{equation}

Substituting with (\ref{Coefficient-Sigma-11-weak}), (\ref{Coefficient-Sigma-21-weak}) and (\ref{Sigma-2-2}) into (\ref{Coefficient-Sigma}) we obtain
\begin{equation}\label{Sigma-1-2}
\Sigma_1=\Sigma_1^{(1)}+\Sigma_1^{(2)}\simeq \frac{\alpha}{15\pi}\left(\frac{eB}{m^2}\right)^2-\frac{4\alpha}{3\pi}\ln\left(\frac{k}{m}\right)
\end{equation}

\begin{equation}\label{Sigma-2-2-2}
\Sigma_2=\Sigma_2^{(1)}+\Sigma_2^{(2)}\simeq \frac{\alpha}{6\pi}\left(\frac{eB}{m^2}\right)^2-\frac{4\alpha}{3\pi}\ln\left(\frac{k}{m}\right)
\end{equation}

Finally, plugging in (\ref{Coefficien-WA}) the results (\ref{Sigma-1-2}) and (\ref{Sigma-2-2-2}) we have

\begin{equation}\label{Coefficien-WA-1-2}
 \Pi_2(eB<m^2<k_\|^2, k_\bot^2)=-\frac{1}{2}\left[\frac{\alpha}{3\pi}\left(\frac{eB}{m^2}\right)^2\left(\frac{k_\bot^2}{5}+\frac{k_\|^2}{2}\right)-\frac{4\alpha}{3\pi}k^2\ln\left(\frac{k}{m}\right)\right]
\end{equation}

\end{document}